\documentclass[sn-mathphys-num]{sn-jnl}%
\usepackage{graphicx}%
\usepackage{multirow}%
\usepackage{amsmath,amssymb,amsfonts}%
\usepackage{amsthm}%
\usepackage{array}%
\usepackage{cellspace}%
\usepackage{changepage}%
\usepackage{framed}%
\usepackage{caption}%
\DeclareCaptionLabelFormat{bold}{\textbf{#1 #2}}
\usepackage{mathrsfs}%
\usepackage[title]{appendix}%
\usepackage{xcolor}%
\usepackage{textcomp}%
\usepackage{manyfoot}%
\usepackage{booktabs}%
\usepackage{algorithmic}
\usepackage[ruled, linesnumbered]{algorithm2e}
\usepackage{listings}%
\newcolumntype{P}[1]{>{\centering\arraybackslash}p{#1}}
\newcolumntype{M}[1]{>{\centering\arraybackslash}m{#1}}

\usepackage{enumitem}%
\usepackage{caption}%
\usepackage{subcaption}%
\usepackage{adjustbox}%

\begin{document}

\title[Beyond Static Datasets: A Behavior-Driven Entity-Specific Simulation to Overcome Data Scarcity and Train Effective Crypto Anti-Money Laundering Models]{Beyond Static Datasets: A Behavior-Driven Entity-Specific Simulation to Overcome Data Scarcity and Train Effective Crypto Anti-Money Laundering Models}

\author*[1]{\fnm{Dinesh Srivasthav} \sur{P} }\email{dineshsrivasthav.p@tcs.com}
\author[2]{\fnm{Manoj} \sur{Apte}}\email{manoj.apte@tcs.com}

\affil[1]{\orgdiv{Cyber Security and Privacy Research}, \orgname{TCS Innovation labs}, \orgaddress{ \city{Hyderabad}, \postcode{500034}, \state{Telangana}, \country{India}}}
\affil[2]{\orgdiv{Data and Decision Sciences Research}, \orgname{TCS Innovation labs}, \orgaddress{\city{Pune}, \postcode{411057}, \state{Maharashtra}, \country{India}}}

\abstract{For different factors/reasons, ranging from inherent characteristics and features providing decentralization, enhanced privacy, ease of transactions, etc., to implied external hardships in enforcing regulations, contradictions in data sharing policies, etc., cryptocurrencies have been severely abused for carrying out numerous malicious and illicit activities including money laundering, darknet transactions, scams, terrorism financing, arm trades. However, money laundering is a key crime to be mitigated to also suspend the movement of funds from other illicit activities. Billions of dollars are annually being laundered. It is getting extremely difficult to identify money laundering in crypto transactions owing to many layering strategies available today, and rapidly evolving tactics, and patterns the launderers use to obfuscate the illicit funds. Many detection methods have been proposed ranging from naive approaches involving complete manual investigation to machine learning models. However, there are very limited datasets available for effectively training machine learning models. Also, the existing datasets are static and class-imbalanced, posing challenges for scalability and suitability to specific scenarios, due to lack of customization to varying requirements. This has been a persistent challenge in literature. In this paper, we propose behavior-embedded entity-specific money laundering-like transaction simulation that helps in generating various transaction types and models the transactions embedding the behavior of several entities observed in this space. The paper discusses the design and architecture of the simulator, a custom dataset we generated using the simulator, and the performance of models trained on this synthetic data in detecting real addresses involved in money laundering.}

\keywords{Crypto forensics, Money laundering, Transaction simulator, Synthetic data, Machine learning}

\maketitle

\section{Introduction}\label{sec1}

Cryptocurrencies \cite{crypto}, a novel form of digital or virtual currency, have been in wide use for quite a few years. These decentralized financial assets leverage cryptography for security and operate on blockchain technology. There are more than a thousand different cryptocurrencies in the market. Of them, Bitcoin is the most used cryptocurrency. Bitcoin holds close to 50\% of the total crypto market share \cite{coinmarketcap}. This is due to many reasons, prominently for its salient properties such as pseudo-anonymity, lack of central authority, ease of cross-border transactions and so on, which are however also the same for many other cryptocurrencies. What makes it widely adoptable, besides it being one of the first cryptocurrencies, is the number of entry, exit options, and third-party services easily available for Bitcoin. Due to a combination of such unique properties offering enhanced privacy and flexibility to make transactions from anyone anywhere to anyone-anywhere without a verified account or source of funds, Bitcoin has also got the interest of cyber criminals who often trade in or bring in illicit funds into the crypto ecosystem through Bitcoin and vice versa. Other privacy-enhanced cryptocurrencies would often be used at the intermediate or layering stages through cross-chain conversion or via Over-The-Counter (OTC) brokers, and so on, for enhanced anonymity, diminishing the traceback chances to the source of funds. This way, cyber criminals are misusing cryptocurrencies to do several illicit activities such as money laundering, ransomware, illicit trading, terrorism financing, and a lot more.\\ 
Of many such crimes, money laundering is a prominent illicit activity taking place through cryptocurrencies as it is often identified that the trails of most of the illicit activities end up with money laundering to integrate them back into the financial ecosystem as clean money so that it can be subsequently used without legal constraints. Last year alone about 23.8 billion dollars were laundered through cryptocurrencies as per Chainalysis crypto crime report \cite{chainblog}. Therefore, identifying money laundering is very crucial and it in turn helps in identifying other illicit activities and prevents major financial losses. And it is also very much needed to detect it and warn the users, banks, and so on, preventing them from making transactions with such illicit entities or reporting the same to law enforcement agencies if identified.
However, the detection of money laundering is not so simple as it has a variety of hardships as discussed below.

\subsection{Challenges of crypto money laundering detection}
There are various types of challenges in this space that may have to be looked through different levels. We summarize them as follows, grouping them into 4 categories.\\
\begin{enumerate}
\item \textbf{Due to inherent features:} 
    \begin{itemize}
    \item (Pseudo-) anonymous accounts on blockchain.
    \item No restriction on the number of addresses/accounts that a single person can hold and create. Though this is not a special case with crypto, the number of accounts that one can create with ease here are unimaginable and cannot be compared with the general bank accounts. For instance, one can have and operate say more than 1000 accounts from the same or different wallets, which is likely not the case with the bank accounts. Besides, there is pseudo-anonymity as mentioned above due to which one would not know that all these accounts are operated by the same entity.
    \item The possibility of the involvement of multiple parties on either side in a single transaction makes it hard to identify who all are the parties involved and how was the fund shared, also in subsequent transactions.\\
    \end{itemize}

\item \textbf{Challenges in customer identification protocols:}
    \begin{itemize}
    \item Many virtual asset service providers (VASPs) do not collect and have customer identification information (such as know your customer (KYC) and customer due diligence (CDD)).
    \item Some VASPs collect minimal information to be considered (semi-) regulated by regulators. However, this information is unreliable and can often be misleading, because launderers usually submit fake details as the VASPs may not verify the authenticity of this information due to lack of standard to verify the same.
    \item Certain VASPs are not willing to share customer or transaction data even with the enforcement agencies.
    \item Since the accounts are anonymous, and when VASPs do not have or share the details, contextual information is the usual way to identify to whom that account belongs to. Contextual information is the information available on the internet - public forums, abuse reports, social media and so on, related to information about an account sometimes revealing its identity, usually given out by the peers of an account with which it has interacted. However, contextual information may not be reliable due to a lack of reliable sources and a lack of validation of available information. For example, a Launderer who has 150 accounts can use 4 of his accounts to review it good on public forums.\\
    \end{itemize}

\item \textbf{Enforcement hardships:}
    \begin{itemize} 
    \item There is no proper mechanism to identify unauthorized or unregulated/ unlicensed exchanges. Such exchanges all being online can con general users into unknowingly getting involved in fraudulent activities or getting away with their funds.
    \item Availability of multiple unregulated options to enter and exit the crypto ecosystem.
    \item Lack of stringent regulations, policies, and non-compliance to those that exist -- Though there are many guidelines put forth by inter-governmental bodies like FATF which stands for Financial Action Task Force \cite{fatf}, many of them are not being enforced -- elaborated further in Section 2. Thereby, becoming a hardship for detection and sometimes a road blocker too. This is also an issue faced by law enforcement agencies, banks, and certain genuine exchange, wallet services, and so on, where the flow of information and verification would not proceed smoothly, or at times, not at all, due to contradicting data sharing policies, or, no data being collected, or, information that has been collected but not verified misguiding investigation, or, creating integrity issues, etc. There are many such open-challenges in this regard.
    \item Funds present as cryptocurrencies cannot be frozen by government, financial institutions, and enforcement agencies. However, bank accounts and other way-outs can be blocked.
    \item The ease of cross-border transactions makes it difficult to get the needed evidence to prove the crime. Funds can be easily transferred to geographies having weak regulations as no entity is there to stop or verify. Due to multiple reasons such as data privacy, lack of KYC, institutional policies, trust issues, varying or contradicting policies and guidelines for different jurisdictions and so on, tracing further, and collection of evidence gets difficult.\\
    \end{itemize}

\item \textbf{Third-party services and tactics:}
    \begin{itemize}
    \item Use of third-party services such as mixers and tumblers to obfuscate the funds -- such services make the transactions and funds tracking process complicated.
    \item Deep integration into the system by using common patterns -- Simple patterns like Coinjoins, use of multiple change addresses, senders and recipients collaborating in the inputs itself and so on, are used, all not just limited to money laundering patterns. These patterns or such transactions are commonly seen on the blockchain in many transactions just for enhanced privacy, which are not related to money laundering. These do not raise any red flags as they seem usual.
    \item Use of privacy-enhanced cryptocurrencies.
    \item The ease of cross-chain conversions (from one cryptocurrency to another cryptocurrency) makes it difficult to track funds.
    \item Presence of virtual and temporary VASPs.
    \item Claims of crypto appreciation owing to the unprecedented fluctuations in the crypto market. This is one of the common money laundering integration methods used to explain the heavy returns which may not be denied due to the volatility in the market.
    \end{itemize}
\end{enumerate}

\section{Related Work}
The Financial Action Task Force (FATF) is an intergovernmental organization established to combat money laundering and terrorism financing on a global scale. It is also often referred to as the watchdog of global money laundering and terrorism financing. Comprising member countries and jurisdictions, FATF sets international standards and develops policies to enhance the effectiveness of Anti-Money Laundering (AML) and Counter-Terrorist Financing (CTF) measures \cite{FATF_recom}. In the context of crypto money laundering, FATF has extended its regulatory purview to encompass VASPs and introduced certain imperative guidelines \cite{fatf2, fatf3, fatf4, fatf6} such as:
\begin{itemize}
	\item Categorization of VASPs: Virtual assets should be defined into one of the existing categories for which there are already certain regulations in place such as into funds, property, currency and so on, so that the existing rules will be applicable to them in addition to the FATF recommendations and guidelines if implemented.
	\item Regulation of VASPs: All virtual asset service providers should be regulated and authorized. There should be some committee or department handling this process of regulating, issuing/canceling licenses and so on.
	\item Know Your Customer (KYC) and Customer Due Diligence (CDD):  All virtual asset service providers should strictly follow KYC and customer due diligence protocols. They need to have and should be able to reproduce the required customer and transaction details when the concerned authority asks about them.
	\item Travel rule: Implementation of travel rule which is about sharing the concerned identifying information of the sender, recipient and transactions for certain or requested fund transfers such as cross-border, and domestic bank transfers involving different exchanges or jurisdictions, transactions carrying over 1000 USD and so on. These measures aim to bring transparency, traceability, and regulatory oversight to the rapidly evolving and decentralized realm of cryptocurrencies.
\end{itemize}

\noindent Apart from the above, FATF has also provided many valuable recommendations, guidelines, compliance assessment measures, and also, has investigated many cases and created many red-flags to be monitored by nations, VASPs, banks, regulatory bodies, etc., \cite{fatf7,fatf8,fatf9,fatf10}. While FATF has indeed introduced the aforementioned to curb money laundering in cryptocurrencies, the persistence of such illicit activities can be attributed to several factors such as:
\begin{itemize}
\item The rapid evolution of the cryptocurrency landscape often outspaces regulatory frameworks. New cryptocurrencies with advanced features and innovative technologies that facilitate blockchain interoperability and make cross-chain conversions easy, with enhanced privacy are constantly emerging, creating loopholes that money launderers exploit before regulations can catch up.
\item Inconsistencies and gaps in enforcement between jurisdictions significantly contribute to the continued money laundering at a global scale \cite{fatf11}.
\item The sheer volume and complexity of recorded transactions involving millions of anonymous accounts make it difficult for authorities to effectively monitor every single transaction.
\end{itemize}

\noindent\newline In literature, many detection methods have been proposed starting from naïve approaches involving manual investigation such as using contextual analysis, with address clustering and tagging heuristics and so on \cite{10.3389/fcomp.2020.600596, 10174963}. But, provably, they are time-consuming and they being simple, already have counter attacks by launderers in place. Most of the subsequent works proposed machine learning approaches to better identify the deep patterns present in numerous layers of money laundering transactions, aiming to reduce the manual effort in investigation \cite{a1,a2,10.1145/3418981.3418984}. However, to use such techniques, we need to have a diversified dataset capturing all possible money laundering patterns to train the model and identify the nature of new transactions. As we know, the blockchain data of cryptocurrencies is ever increasing (for instance, Bitcoin blockchain's data in its current size is over 500 GB). Thus, analyzing such amount of data from its raw format, identifying labeled accounts from it, capturing their transactions are all heavily computational and resource-consuming activities needing a multi-core system with hundred's of gigabytes of RAM, many terabytes of storage and processing space, and many other connected distributed nodes working together \cite{graphsense} which is pretty expensive.  This is also a major persistent problem observed in the literature.\\
To address this, they have trained models basis a dataset curated for a very concise scope or a small dataset captured over a short interval. However, these often lacked diversity, labels, balanced classes, scalability, customization, and so on. One such widely used or cited dataset is the Elliptic dataset \cite{weber2019antimoney} which was developed by a crypto analytics firm, Elliptic. It is a time-series graph dataset that encouraged the application of temporal analysis, and graph-based methods such as graph convolutional network (GCN), graph neural network (GNN), and so on. Despite its significant contribution to advancing analytics in this space, again, this is a static graph dataset captured just over 2 weeks, having about 2 lakh nodes. It has a serious problem with data annotation. Out of all the samples, only 23\% are labeled which is only 46 thousand samples, and the rest are unknown leaving an ambiguity. Also, of this 23\%, only 2\% is illicit and the other 21\% is licit posing a severe class imbalance problem as well.\\
Apart from this, techniques like active learning have been proposed \cite{10.1145/3383455.3422549, make5040084}, where we increase the size of the dataset with the model's prediction on new samples. There has been an identified challenge \cite{DASGUPTA20111767} with this approach specifically for this scenario, which is the model would initially be trained on a small dataset leading to overfitting or bias issues, and when we use the same model for predicting new samples and adding them to dataset, the ground truth may not the be the real truth and can lead to wrong predictions subsequently.\\

\hspace{-0.6cm} A summary of some of the key limitations of the current technology:
\\
\begin{enumerate}
    \item \textbf{Manual data collection:}
\begin{itemize}
\item This is strenuous because the Bitcoin core node has data in a raw format and is mammoth in size, thus is difficult to operate. 
\item As transactions span across a long timeframe, one needs to manage multiple dumps, if using data dumps, and extract the needed transactions which is a laborious job. 
\item There are API limitations if using a third-party block explorer.\\
\end{itemize}

\item \textbf{Dataset related:}
\begin{itemize}
\item There are not many good labeled datasets.
\item No scalable or customizable (that is, static) dataset is available. 
\item Entity-level categorization is missing in existing datasets since they often classify data into two high-level classes of licit and illicit without proper analysis of the underlying category.
\item Small datasets present have severe class imbalance problems with lack of needed distribution for specific use cases.\\
\end{itemize}

\item \textbf{Data generation related:}
\begin{itemize}
\item Patterns change over time. No dynamic way to add more transactions or samples with modified patterns.  

\item Data generation models such as variational autoencoder (VAE), generative adversarial network (GAN), and so on, were also proposed \cite{gan} to generate synthetic data close to real-world transactions. Nonetheless, these techniques would need the training samples to fully represent the intended data and the patterns required at least to some extent to be able to generate all sorts of data needed. However, this is the major setback we have, owing to the pseudo-anonymity of crypto accounts and their transactions. Besides, generative models like GANs not only require mammoth quality training data with rich entity behavior and patterns, but also require great infrastructure capabilities and a lot of time to train and fine-tune them to obtain the best possible results.

\end{itemize}
\end{enumerate}
To address these major problems and to achieve a detection model with limited hardware, we came up with a methodology as discussed in the next section(-3).

\section{Methodology}
Based on the findings from the literature, it is evident that we need a method to obtain the transactional data of specific accounts of interest or even a chain of transactions of their neighbor accounts, basis the very purpose of investigation or use case building, with feasibility to customize the occurrence of specific patterns or entities. This helps us in having a scalable, customizable, labeled, entity-specific dataset. This is what we exactly did. We developed a Behaviour-embedded entity-specific Bitcoin-like money laundering transaction simulator (can also be applied to any other cryptocurrency following the Unspent Transaction Output (UTXO) mechanism) that helps in generating patterned transactions in needed requirements, addressing most of the limitations identified in the literature.\\
The information about the behavior of different entities as to how they interact and function is vital for building such a simulator. However, this is not available. Therefore, we started with an exhaustive explorative investigation, tracking and identifying the same. This investigation was aimed to capture every minute detail possible corresponding to the behavior of various entities that are often observed in this space (described in Section 3.1.1). We began with tracking transactions of illicit accounts that were identified to be involved in money laundering as reported by governmental bodies such as OFAC (Office of Foreign Assets Control) through its specially designated nationals and blocked persons list \cite{sdn}, U.S. Department of Justice through its press releases \cite{doj}, and so on, which covered the Bitcoin/crypto addresses associated with many money laundering and illicit cases. The chain of transactions of both the reported accounts as well as the transactions of their neighbor accounts were tracked and analyzed for patterns representing their mode of operation. Additionally, several case studies involving various scenarios and methods of money laundering were studied from several sources including public releases by inter-governmental bodies, law enforcement agencies, regulatory red-flags, crimes reported, and so on. In addition to this, to identify the behavior of named entities such as exchanges, OTC brokers or nested exchanges, money mules, etc., transactions and interactions of labeled accounts available in the public domain \cite{reports1, harvard} were explored in association with entity-mapping services or repositories \cite{reports2} and abuse reports \cite{reports3,reports4,reports5,reports6}. Many third-party services were also explored and tried including mixing services, decentralized exchanges, escrow services, and so on, to track the flow of funds, understand their functionalities, services available, fees involved, minimal requirements for transaction making, additional charges for special requests such as faster transfer, use of more number of addresses in transactions, increasing the number of transactions to reach the destination address, etc.\\
Through this exhaustive explorative study, we, therefore have captured many precise details such as what kind of entities have they interacted with, how was the money split at various steps, the number of accounts used by different entities for different splits, time gap between their transactions, the scale of money being transacted each time, transaction fees spent and patterns in it, how do they involve licit accounts in their flows, how money mules come into picture, dusting attacks, how proxy accounts are used, how often are they revised, how do mixers mix the funds, what type of accounts do they use,  how is this fund integrated back as legit money, and so on, for different entities and their proxy accounts.\\

\subsection{Design and architecture of the Simulator}
Using this foundational domain knowledge, we constructed the simulator as a collection of 29 modules, where 3 modules are to check the availability of an account for a specific type of transaction, and 4 modules are to compute and update the transactional values for different transaction types, 5 modules are for generating entity specific transactions, 1 module to generate pattern and entity agnostic transactions, and the rest are transaction generating modules each acting as a template to generate transactions carrying certain specific behavior related to an entity or transaction patterns. The algorithm for the same is discussed in Section 4.\\
 
Figure 1 is the block diagram of the simulator. There are three blocks depicted in the diagram: input specification, transaction simulator, and simulated transactions. Input specification describes the input parameters needed for simulation. They may either be directly passed to the respective transaction generation modules or can be consolidated as a schema or a mindmap which can be transformed into an executable for transaction simulation. The input parameters needed are: Address type – sender and receiver address types (and quantities) (description of address types (or) entities is given in Section 3.1.1),   type of transaction if any specific pattern is required, quantity of transaction set, and duration for the same. These four parameters are the input arguments for the 17 transaction-generating modules as mentioned earlier. For the 5 modules that generate entity-specific transactions, we require only the quantity of respective transactions needed, as we have templates created for ready-use for a set of entities. Section 4 discusses more on this part.\\
The second block is the transaction simulator, which takes in the input specification and yields simulated transactions. A simulator is a collection of different transaction generation modules each concentrated towards generating different types of patterns related to various entities and transaction types. Each entity typically follows certain behavior or transaction types and sometimes a combination or a sequence of a few. The internal blocks of the transaction simulator depict this relation. The simulator generates about 14 (outer layer accounts are considered just for the sake of completion, further elaborated in Section 3.1.1) types of entities’ behaviors. The ‘entity types’ block depicts them. Similarly, there are different transaction types it generates based on the input specifications and requirements which are broadly given in the ‘transaction types’ block ('transaction types' are described in Section 3.1.2).\\
The third block, simulated transactions, are the output transaction sets generated based on the input specifications given, which consists of different pattern-embedded transactions which can also include the regular transactions, all of which are simulated using the simulator.\\

There are numerous types of accounts (based on what and how they operate) and many types of transactions as identified on the Bitcoin blockchain or any other blockchain for that matter. However, delving into the specific scenario of money laundering, we identified that there are a set of entity and transaction types that appear often in many trails, as described in Sections 3.1.1 and 3.1.2. These are also the entity and transaction types we considered for the construction of the simulator as depicted in Figure 1.\\

\begin{figure*}[ht!]
\setlength{\fboxsep}{10pt}
\fbox{\includegraphics[width=\textwidth]{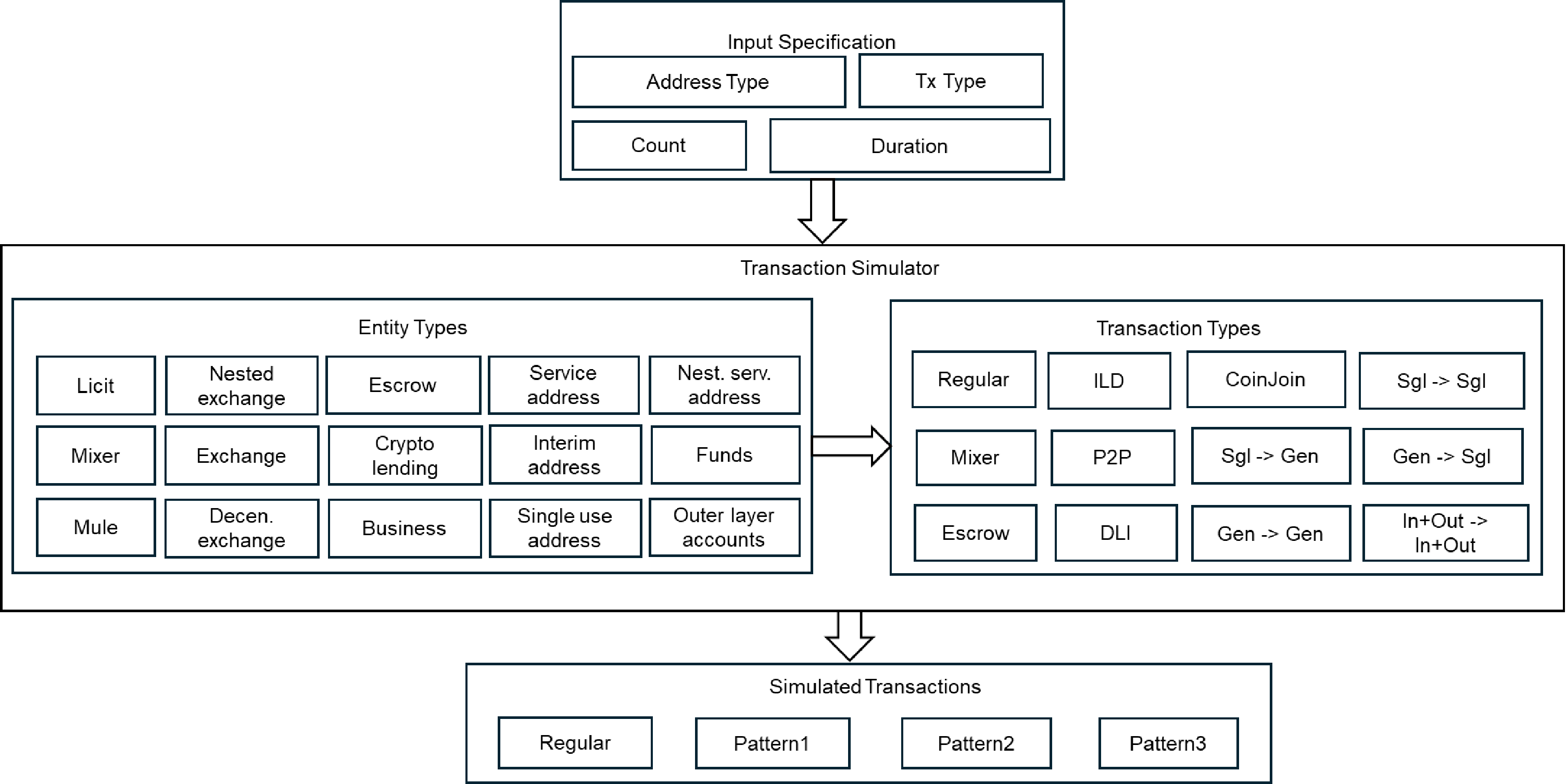}}
\caption{Block diagram of the simulator}
\label{fig:figure1}
\end{figure*}

\subsubsection{Description of entities}
\begin{enumerate}
\item Licit: These are genuine accounts usually related to general Bitcoin or crypto users, miners, verifiable organizations, and so on. These are not involved in illicit activities. However, certain illicit accounts may occasionally send certain small funds to them as a layering technique in money laundering. 

\item Exchange: Crypto exchanges play an important role in circulating Bitcoins. Exchanges are usually centralized where they own a large number of accounts in Bitcoin and across other chains and platforms as well. They take fiat money off-chain and transfer Bitcoins to their customers who thereafter make transactions with them. Similarly, they also help in transferring Bitcoin between different Bitcoin wallets and accounts and help in converting Bitcoin back to fiat money or to altcoins.  

\item Decentralized exchange: These do the same as Exchanges with a difference in the way they operate. Unlike exchanges, these are decentralized. They usually do not own a large number of accounts and are also not very widely used compared to exchanges. Decentralized exchanges use an escrow mechanism to facilitate the transfer of funds. For instance, if A wants to exchange something with B in a decentralized way, then A and B can transfer their funds to an escrow account which is a multi-signature wallet that needs at least two signatures to unlock the funds. A and B with their keys will have to sign the transactions approving the transfer of funds as intended from the escrow account. The funds won’t be transferred if either party is not signing the transaction. In case of a dispute, the decentralized exchange can use its key to sign the transaction upon verification of the concern. Security deposits will have to be deposited from both parties in the beginning to facilitate dispute resolution.    

\item Nested exchange: These are not regular exchanges. They have their accounts in exchanges but are not exchanges themselves. They use a chain of transactions to transfer the funds across different accounts in different exchanges and wallets. These are usually used as a layering technique. 

\item Escrow: Described in decentralized exchange. 

\item Mixer: These are third-party services available in the crypto ecosystem that facilitate the mixing of funds from different parties thereby making it difficult to figure out who is sending/receiving from whom and what is being transferred. They follow a complicated trail of transactions to do this by mixing funds from various accounts. 

\item Mule: These are considered licit. However, knowingly, or unknowingly, they get involved with illicit activities where they make certain transactions from the large illicit trail. 

\item Funds: These are money reserves that entities such as exchanges, and mixers would use to exchange a currency or facilitate mixing. Similar accounts are also deployed by illicit parties to fund their transactions.  

\item Business: These are accounts used by illicit entities showcasing them to be related to certain legitimate business activity. However, they are used in the integration phase of money laundering to show a legititimate source for laundered money. 

\item Crypto lending: This is also a type of business category where they create a crypto lending platform. Retail borrowers deposit their crypto assets in the platform for a certain period to get fiat cash. The platform will use this deposited crypto to invest in other businesses or offer services. Such businesses or customers will pay the platform a certain amount periodically which will be shown as a legitimate source of their illicit funds. We have two parties here: the lending platform and investors. 

\item Service address: Exchanges employ a large set of addresses usually referred to as service addresses who make transactions on their behalf in dealing with different customers. Service addresses are revised frequently. 

\item Nested service address: Similar to service addresses but deployed by nested exchanges. These predominantly interact with service addresses of different exchanges to gather funds to transfer. They extended the transaction chain length for the nested exchange. 

\item Interim address: These are the accounts used by illicit entities to make their transactions and interactions with different entities like mixers, exchanges and so on. 

\item Single-use address: These are one-time-use addresses. They receive once and they send once, and they are never used again. These are widely used by all the entities to make transfers and extend the chain length, so that, in case, an entity they dealt with is found to be illicit, they can be safeguarded as they are a few hops away from the illicitly identified party due to transactions made with single-use addresses.

\item Outer layer accounts: Transactions are a continuous chain of interactions between numerous accounts and therefore, it is not practically possible to simulate infinite chains. Hence, one would usually scope this to a certain limit such as accounts of interest and their immediate neighbors, etc. In this case, this is 2 levels or 1 hop of transactions where we are considering level 1 as addresses of our interest, and level 2 as their neighbors, which is 1 hop from the interested accounts (interested accounts $\leftrightarrow$ neighbors). We can extend this to 'n' hops where we consider the neighbors of 1st level accounts in 2nd level and neighbors of 2nd level accounts in third level and so on.\\
Consider we need 1 hop of transactions which is A$\leftrightarrow$B($\leftrightarrow$C). 'A' is a set of one or more accounts of our primary interest which can include addresses of different categories. 'A' has its transactions with 'B' which are its neighbors. Since we want 1 hop of transactions, we are interested in the behavior of 'A' and 'B'. As we are simulating all the interactions of 'A' with its immediate neighbors (that is, 'B'), we get the complete behavior of 'A'. However, to complete the behavior of 'B', which is our second level, we have to also consider their transactions which are done with their neighbors in the third level, which are not considered above in our interest. For such a simulation, we use a set of accounts that make transactions with our last level of accounts considered, and we are not considering this set of accounts in our scope. In this example, our scope is 'A' and its neighbors which are 'B'. To complete behavior of 'B', we consider transactions of 'B' with their neighbors 'C' which fall in level 3. As level 3 is not part of the considered scope, we do not simulate further. Thus, here, 'C' is the outer layer accounts which are just used to complete the behavior of 'B' but are not in our interest.
\end{enumerate}

\subsubsection{Description of transaction types}
Certain transaction types like 'Mixer', 'Escrow', go with the name of their concerned entity as described below.
\begin{enumerate}
\item Regular: These are the usual transactions; they typically do not follow a certain pattern and are the widely seen transaction type.

\item DLI: Denotes chain of transactions from Depositor $\rightarrow$ Lender $\rightarrow$ Investor. Used to make the mentioned chain of transactions between three parties and helps to track the funds deposited by respective accounts of depositors, funds retained by the lending platform, and funds transferred to respective investor accounts from the respective lender accounts.   

\item ILD: Denotes chain of transactions from Investor $\rightarrow$ Lender $\rightarrow$ Depositor. Also, used to compute and return the interest for the fund deposited by the respective lender accounts and in turn return an equivalent to the depositor accounts. 

\item P2P: Represents peer-to-peer transactions. Used for transferring funds from party1 to escrow, party2 to escrow, security depositing, and escrow to party1 and party2 deducting a certain \% as platform fee, for an overall transfer from peer1 (party1) to peer2 (party2).  

\item Escrow: Subclass of P2P to track and facilitate the exchange of deposited funds, transfer back of security deposits, deduction of platform fee and sending this fee to the account of the decentralized exchange. 

\item Mixer: This has four sub-classes to represent different types of mixers. Each of those facilitates mixing in different ways. 
\begin{enumerate}
    \item Sub-class-1: A sequential combination of 5 different types of transactions is called with the intervention of 'funds' at step-3. The number, sequence of module invokes, and the invoked modules here are different from other mixers implemented. This is true for the remaining sub-classes as well.  
    \item Sub-class-2: A sequential combination of 5 different types of transactions is called with the intervention of 'funds' at step-3. Single-use addresses and Coinjoin-like transactions are predominantly used here. 
    \item Sub-class-3: A sequential combination of 13 different types of transactions is called with the intervention of 'funds' at step-4, 6, and 9. Single-use addresses and Coinjoin-like transactions are predominantly used here.
    \item Sub-class-4: A sequential combination of 9 different types of transactions is called with the intervention of 'funds' at step-3, 5, and 7. Coinjoins and Coinjoin-like transactions are predominantly used here.
\end{enumerate}

\item Coinjoin: It is a transaction type where both sender and receiver collaborate and act as the senders and receivers of the transaction (Sender+Receiver $\rightarrow$ Sender+Receiver). Every account receives an equal amount thereby making it difficult to understand who is transferring to whom, and trace the subsequent transactions.  

\item Sgl $\rightarrow$ Sgl: Represents transactions between two sets of single-use addresses. Helps to track the occurrence of a single-use address and remove it if its limit is exceeded. Sender single-use addresses cannot send any further and receiver single-use addresses cannot receive any further. There is an additional subclass to facilitate the coinjoin type of transaction along with Sgl $\rightarrow$ Sgl behavior to additionally incorporate the same-value-transfer property. 

\item Sgl $\rightarrow$ Gen: Represents transfer from a set of single-use addresses to a set of general addresses. Sender single-use addresses cannot send any further. There is an additional subclass to facilitate the coinjoin type of transaction along with Sgl $\rightarrow$ Gen behavior to additionally incorporate the same-value-transfer property. 

\item Gen $\rightarrow$ Sgl: Represents transfer from a set of general addresses to a set of single-use addresses. The recipient single-use addresses cannot receive any further. There is an additional subclass to facilitate the coinjoin type of transaction along with Gen $\rightarrow$ Sgl behavior to additionally incorporate the same-value-transfer property. 

\item Gen $\rightarrow$ Gen: Represents transfer from a set of general addresses to a set of general addresses. There are certain patterns seen here such as having only a specific number of accounts on one side or either side of a transaction. 

\item In+Out $\rightarrow$ In+Out: Senders and receivers of a transaction collaborate and act as senders and receivers of a transaction like in Coinjoin but here, it is not needed that the amount received from any account needs to be the same. It is used to make transactions between different entities combining them on either side. 
\end{enumerate}

\section{Process of simulation}
Since the simulator is concentrated on generating money laundering scenarios, we consider transaction generation to be a sequential generation of the flow of transactions where a set of transactions will be carried forward by the subsequent or later steps that result in a chain of interactions between different entities or accounts (addresses). Henceforth, to simulate transactions to specific requirements, we would need to know the same, that is, what type of entities are needed, how should they be linked, transaction flow, or if any sequence of patterns is required, the number (quantity) of transactions needed, preferred timestamp for transactions (in what period should the transactions be generated). A transaction schema or a mindmap can specify such information which is essentially the requirement for generating transactions.\\

\begin{figure*}[ht!]
\centering
\fbox{\includegraphics[scale=0.6]{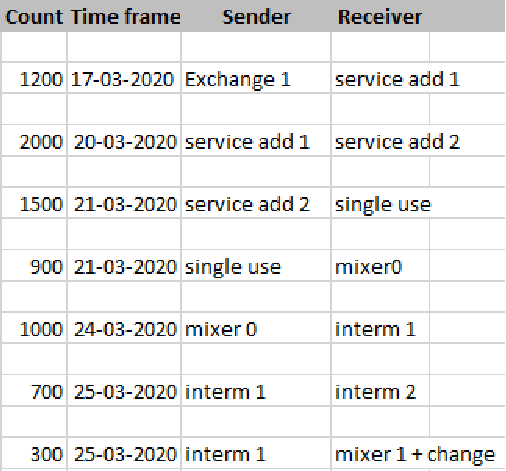}}
\caption{Sample of transaction schema}
\label{fig: figure2}
\end{figure*}

This schema can be provided through an Excel, Comma-separated Values (CSV), JavaScript Object Notation (JSON), or any other similar format with the following information and structure:
\begin{enumerate}
\item If you want to generate transaction(s) from say, A to B, then a row in Excel should describe four attributes of this transaction(s): A, B, number of such transactions needed (that is, from A to B), a time frame where these should preferably take place.\\ That is, count of a specific set of transactions, it could be 1 or more. Another parameter is any desired timestamp for the respective set of transactions which would be useful if someone is doing a timeseries analysis, matching against some activity or an event, and so on. The other two parameters are the sender and receiver as to which entity we want to have as the sender and recipient of the respective set of transactions. This helps in customizing the interactions or flow of funds between different entities as required for varying needs. 
\item The subsequent rows can either continue this chain or have a disjoint transaction set. This same information can be repeated for as many sets of transactions as needed. 
\end{enumerate}

\begin{figure*}[ht!]
\fbox{\includegraphics[width=\textwidth]{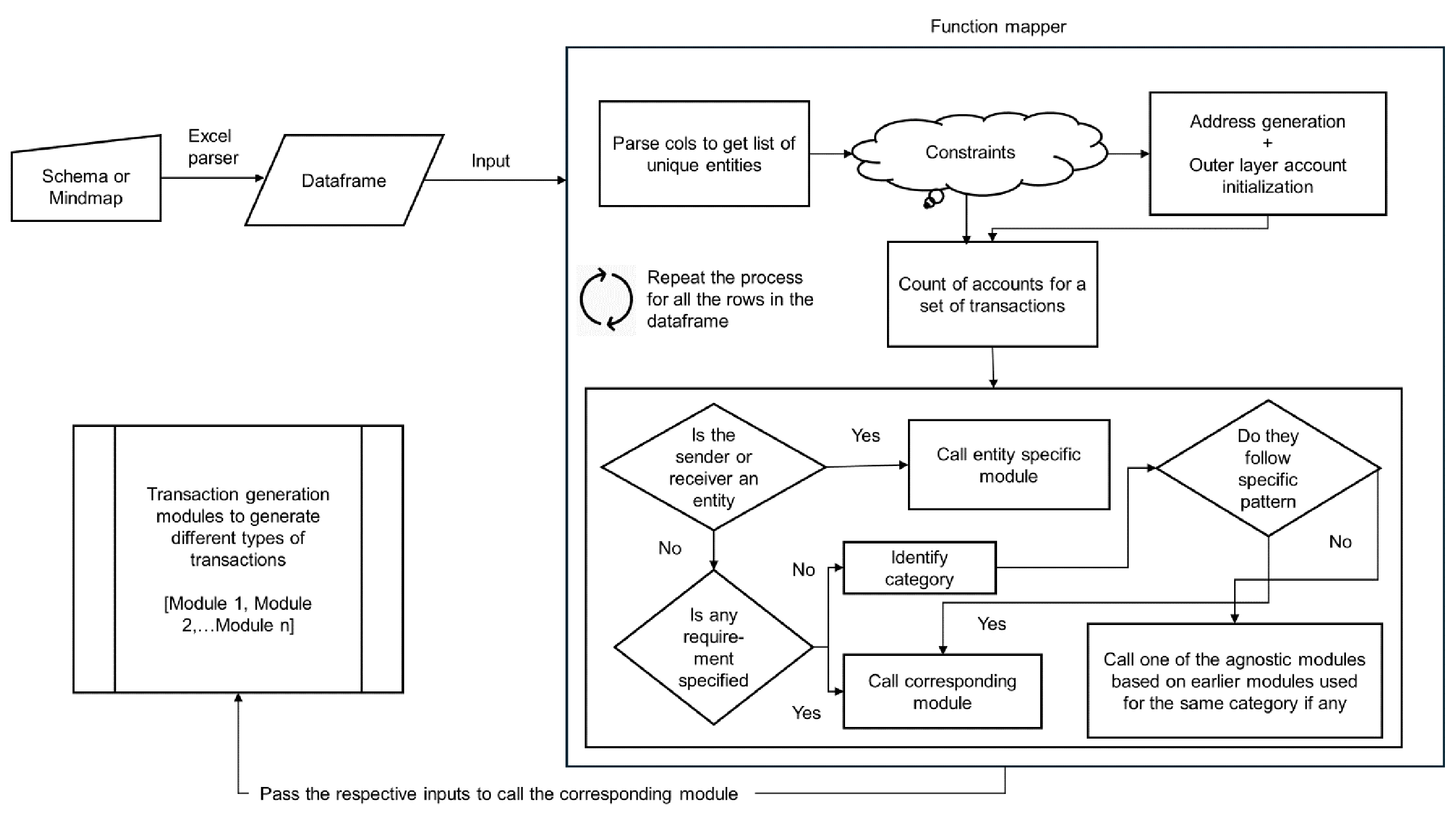}}
\caption{Working of the function mapper module}
\label{fig:figure3}
\end{figure*}

A sample of the schema is given in Figure 2 for reference. Here, we have 1200 transactions from Exchange 1 to Service address 1 in the timeframe of 17th Mar, 2020 (timeframe here is a period between two instances. We take instance-1 as the latest timestamp of the last transaction of all the accounts involved in a transaction. Instance-2 is the date mentioned in the schema like 17/03/20, here. This is further discussed in the later parts of this section), and the next set of transactions are from service add set-1 to set-2 with a quantity of 2000 and in the timeframe of 20th Mar, 2020 and so on. Here, there will be different accounts created for each of these entities just like it happens in real. And, we can maintain a flow of funds between transaction sets if desired like in here, we see funds move from exchange to service address set-1 and then to set-2. Set-1, set-2, and so on (represented with only a number without the word 'set' in Figure 2) are the instances of different entities (For example, service add 1 and 2 represent two disjoint sets of service addresses).\\

\begin{algorithm}[H]
\SetAlgoLined
\SetKwFunction{avail}{avail}

\KwIn{senders (list of sender accounts)}
\KwOut{ac (availability of sender accounts)}

\BlankLine
\SetKwProg{Fn}{Function}{:}{}

\Fn{\avail{senders}}{
    Initialize lists \textit{avail} and \textit{ac}\;
    
    \ForEach{sender in senders}{
        Extract the list of UTXOs for the current sender from the global variable \textit{utxof}\;
        
        \If{the list of UTXOs is not empty and the first element of the list is an integer or a float}{
            Find the maximum value in the list of UTXOs\;
            
            \If{the maximum value is greater than 8000}{
                    Add a sublist to \textit{avail} containing the sender repeated a number of times equal to the count of UTXOs with values greater than 8000\; 
                    
                    //where 8000 is the threshold set to avoid dusting attacks of txn value 5460 sats + fee
            }
        }
    }
    Flatten the list \textit{avail} into a single list \textit{ac}\;
   
    \KwRet{ac}\;
}
\caption{Pattern-agnostic Availability Check Function}
\end{algorithm}

\begin{algorithm}[H]
\SetAlgoLined
\SetKwFunction{txn}{txn}
\SetKwFunction{avail}{avail}
\SetKwFunction{update}{update}
\SetKwProg{try}{try}{:}{}
\SetKwProg{catch}{catch}{:}{end}

\KwIn{var1 (senders), var2 (receivers), var3 (quantity), mini, ts}
\KwOut{None (Generates transactions)}

\BlankLine
\SetKwProg{Fn}{Function}{:}{}

\Fn{\txn{var1, var2, var3, mini, ts}}{
    Initialize lists or variables for invalue, inputs, ind, fees, m, and outputs\;
    Global variables hashes, last\_time, utxof, previous\_state\;
    
    \tcp{Attempt to execute the transaction}
    \try{}
    {
        ac = avail(var1)\;
        
        \For{l in range(var3)}{
            Choose a subset of sender accounts (inputs) with sample size $\geq$ \textit{mini}\;
            
            Calculate the total input value (invalue) from the chosen inputs\;
            
            //Input value of a sender account will be from one of its available UTXOs
            
            Remove the input values from the sender accounts's UTXO (utxof) and update the available accounts list (ac)\;
            
            Calculate transaction fees based on the number of inputs\;
            
            Calculate the number of outputs (m) based on the remaining value after deducting fees\;
            
            Choose receiver accounts (outputs) from var2 based on the number of outputs computed\;
            
            \tcp{Update the transaction records and available accounts}
            Invoke \textit{update} function
        }
    }
    \tcp{Revert to the previous state in case of failure}
    \catch{Exception}{
        Revert to the previous state\;
    }
}
\caption{Pattern-agnostic Transaction Generation Function}
\end{algorithm}

\begin{algorithm}[htbp]
\SetAlgoLined
\SetKwFunction{update}{update}

\KwIn{inputs, outputs, in\_value, fees, ts, ac, var3}
\KwOut{None (Updates records)}

\BlankLine
\SetKwProg{Fn}{Function}{:}{}

\Fn{\update{inputs, outputs, in\_value, fees, ts, ac, var3 (quantity)}}{
    Calculate the remaining value (rem) after deducting fees and output values\;
    
    Calculate the value for each output (out\_value) based on the remaining value and the number of outputs\;
    
    Generate a unique hash for the transaction\;
    
    Update the timestamp for the transaction\;
    
    Update the last timestamp for the involved accounts\;
    
    Update the UTXOs for the receiver accounts\;
    
    Update the database with generated txns\;
}
\caption{Pattern-agnostic Update Function}
\end{algorithm}

Algorithm to simulate transactions given a transaction schema:
\begin{enumerate}
\item Receiving a transaction schema comprising a description of sets of transactions, wherein each set of transactions comprises one or more senders, one or more receivers, the number or quantity of transactions to be generated, and the latest preferred timestamp for the transactions. 

\item Transforming the transaction schema into a data frame using a schema parser (For example, an Excel parser when schema is provided as an Excel). 

\item Generating an executable from the data frame using a function mapper (described in Figure 3) as follows by: 
\begin{enumerate}
\item Parsing the data frame, to identify unique entities in the sets of transactions and transaction type of each set in the sets of transactions. 

\item Creating sets of addresses for the entities, based on a set of factors that help in identifying the requirement of a certain entity. These set of factors (constraints in Figure 3) include:
\begin{enumerate}
\item Number of times each of the one or more unique entities are described in the transaction schema; 
\item Whether each of the one or more unique entities is part of the one or more inputs or the one or more outputs; and 
\item The quantity of one or more sets of transactions corresponding to each of the one or more unique entities in an associated set of crypto transactions, is described in the transaction schema. 
\end{enumerate}
\begin{figure*}[ht!]
\fbox{\includegraphics[width=\textwidth]{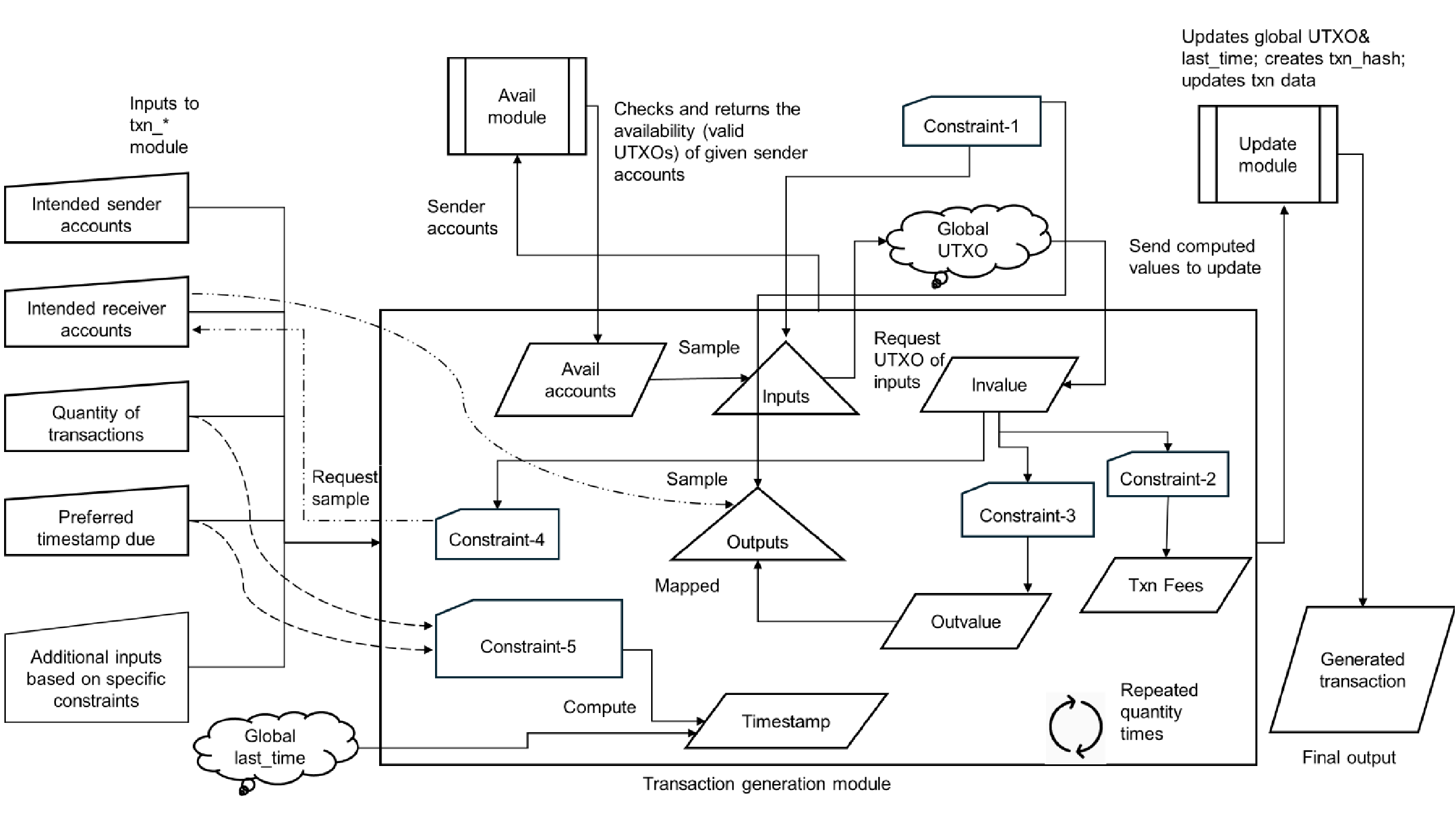}}
\caption{Working of transaction generation module}
\label{fig:figure4}
\end{figure*}
\item Initializing the outer layer accounts and thereby entities, such that the entities have certain existence obtained from transactions before the oldest timestamp specified in the dataframe/transaction schema. 

\item For every transaction set in the schema,  
\begin{enumerate}
\item Selecting one or more appropriate entity accounts corresponding to the sender and receiver respectively. 

\item Mapping each set of transactions to the respective transaction generation module based on a sequence of checks as described in Figure 3, to obtain an executable file for generating the transaction dataset. 
\end{enumerate}
\end{enumerate}

\item The obtained executable would have a sequence of invokes to the respective transaction generation modules, which upon executing, generates the required transactions. The following steps happen within a transaction generation module where the transactions for each transaction set defined in the schema will be simulated by performing the following steps iteratively, wherein the number of iterations is equal to the number of transactions to be generated (the following is for generating a typical transaction or that has no specific constraints; each of the transaction generation modules has their deviations (in one or more of the below steps) from this process to embed specific behavior or pattern in the corresponding transactions):

\begin{enumerate}
\item Checking the availability of one or more accounts of the sender to identify one or more available accounts for performing the transaction. -- Availability specifies if the given account has a UTXO and is sufficiently large (threshold criteria basis transaction type) to accommodate one or more recipients along with the transaction fees for the transaction. It also specifies how many such UTXOs are there for a given account. 

\item Taking a sample of the obtained available accounts to get sender accounts.

\item Obtaining UTXOs of each of the one or more sender accounts and computing InValue as the sum of the UTXOs.  

\item Deducting transaction fee from the InValue to obtain net InValue amount. 

\item Identifying the number of receiver accounts the sender accounts can accommodate based on the net InValue amount. 
\item Distributing the net InValue amount among the identified receiver accounts based on the transaction type and any associated constraints to get the Outvalue. 

\item Computing the timestamp of the transaction. 

\item Assigning a unique alphanumeric transaction hash to the transaction.  

\item Recording the unique alphanumeric transaction hash, sender accounts of the transaction, receiver accounts of the transaction, UTXO put in by the respective sender accounts (InValue), OutValue received by the respective receiver accounts, timestamp of the transaction, transaction fees, in the transaction dataset. 

\item Adding received amounts by receiver accounts to their respective UTXOs.
\item Eliminating UTXOs spent by sender accounts from their respective UTXOs.  
\end{enumerate}

\begin{figure*}[ht!]
\fbox{\includegraphics[width=\textwidth]{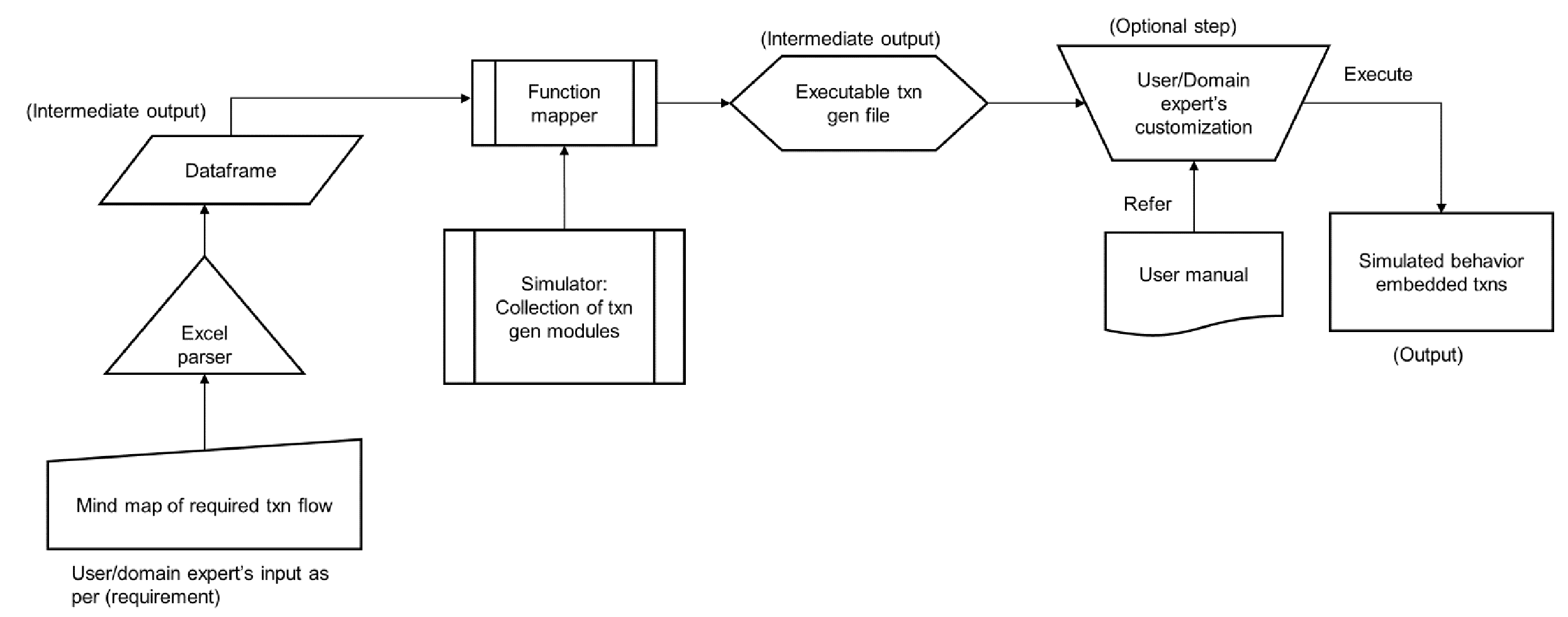}}
\caption{Process flow to use the simulator}
\label{fig:figure5}
\end{figure*}

\noindent\newline Figure 4 demonstrates the above process of pattern-agnostic transaction generation, and algorithms-1, 2, and 3 respectively represent the pseudocode of the functioning of the avail (4a above), transaction generation (4b-4e, 4k above), and update (4f-4j above) modules for this case. The constraints represented in the figure(-4) are to embed concerned entity-behavior that deviates from the agnostic generation process stated above. Based on these constraints, various transaction generation modules were accordingly designed for the simulation of different transaction patterns and entity-behaviors, as mentioned earlier. These constraints are as follows:
\begin{enumerate}
\item Based on the nature of inputs/outputs (For example, if it is a single-use account, there will have to be specific changes, such as suitable measures to ensure that this account is not acting as either inputs or outputs more than once, and so on).
\item Computation of transaction fees based on programmed constraints in cases such as escrow, decentralized exchange, etc., where there are fixed trading limits, and hence, fee changes accordingly. For other cases, it is computed adaptively based on the respective transaction using the equation:

\vspace{0.1cm}
$fees = len(inputs)*1810+0.0008*(sum(in\_value)-1810-5460)+100$
\vspace{0.15cm}
\item Constraint on distribution and processing of invalue to obtain corresponding outvalue. This is different for different modules such as ILD, P2P, Coinjoin and so on (For example, for Coinjoins, all the accounts receive the same outvalue; for ILD, the amount received should be proportionate to the amount invested, and so on).
\item The number of recipient accounts can the inputs accommodate. This is computed adaptively based on the invalue and the transaction fees for the corresponding transaction using the below equation. Here, '5460' is the minimum value an account should transfer/receive to avoid dusting attack.

\vspace{0.1cm}
 $m = int((sum(in\_value)-fees)/5460)$
\vspace{0.15cm}
\item The timestamp is calculated using Uniform or Gaussian distributions (based on cases) in the interval of [max(last\_time of all accounts involved), given timestamp], with sample count as the quantity of transactions.\\

\end{enumerate}
\end{enumerate}

Additionally, 5 modules (entity-specific transaction generating modules as mentioned in Section 3.1)  were designed to explicitly simulate a generalized behavior of a few specific entities namely Mixers, Exchanges, P2P transactions using Escrow, Nested exchanges, and Licit or general transactions. These modules do not need any input except the quantity of transactions required. They do all the above steps with an explicit simulation of the aforementioned entities, which, therefore, can be used without any prior or subsequent transaction flow or UTXO requirement.  They will auto-initiate the accounts, their UTXOs, and timestamps, and call the corresponding transaction generation modules, such as entity-agnostic, and modules corresponding to a few commonly used transaction types, with their input parameters to generate 'regular' type of transactions, and so on. These modules are focused on ready utility without having to prepare a transaction schema. However, for specific requirements, the other 17 transaction-generating modules can be utilized.\\
Figure 5 demonstrates the overall process flow to use the simulator.

\section{Data generation}
Using the constructed behavior embedded entity-specific Bitcoin-like transaction simulator, we have generated about 2 lakh transactions affiliated with about 1.2 lakh accounts of various entities. Figure 6 displays the proportion of the entities used for this simulation, while, figure 7 depicts the class mapping (that is, the proportion of accounts from each category mapped to licit or illicit category). 
\begin{figure*}[ht!]
\centering
\fbox{\includegraphics[scale=0.45]{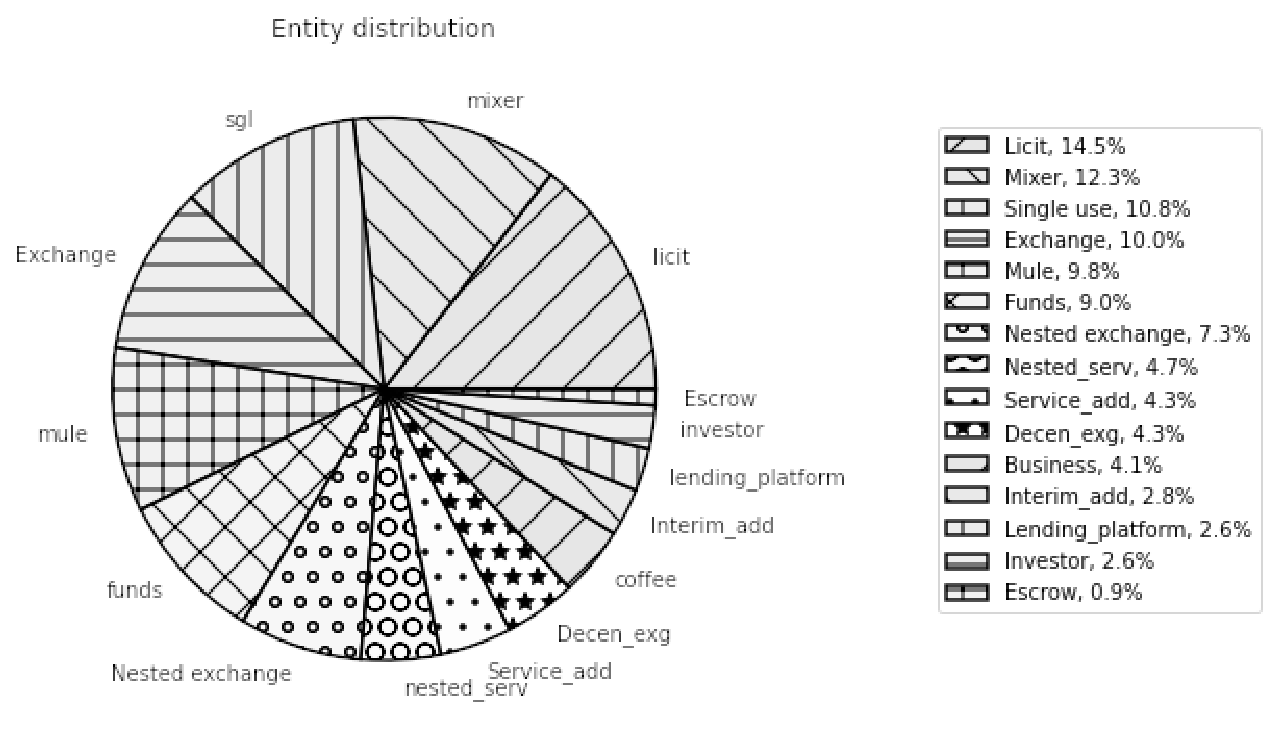}}
\caption{Entity distribution in synthetic dataset}
\label{fig:figure6}
\end{figure*}
\begin{figure*}[ht!]
\centering
\fbox{\includegraphics[scale=0.45]{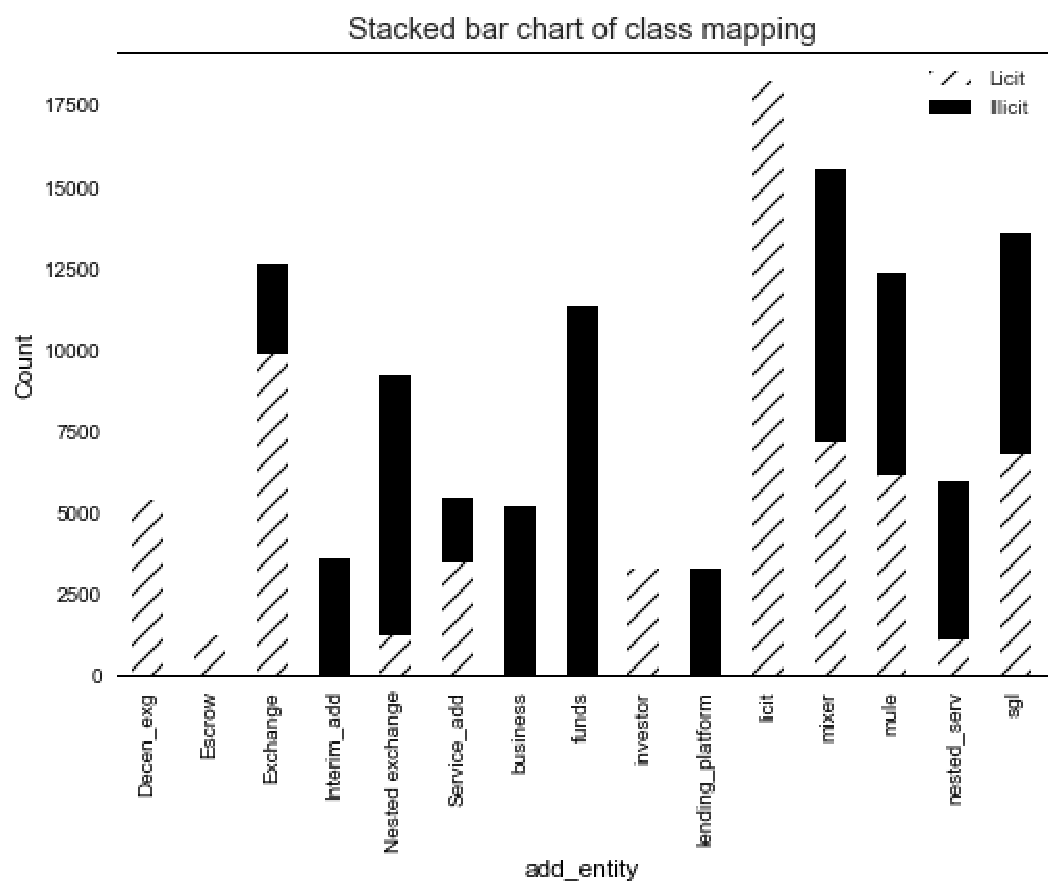}}
\caption{Stacked bar chart of class mapping in synthetic dataset}
\label{fig:figure7}
\end{figure*}
The transaction schema prepared for this simulation represents an end-to-end on-chain Bitcoin money laundering scenario including the three typical phases of money laundering: placement, layering, and integration. The transaction schema consisted of about 132 different sets of transactions representing the said process, with approaches as depicted in Table 1. These 132 different sets of transactions, the way different entities are interacting has been represented in Figures 8, 9, where Figure 8 showcases entity-instance-specific interactions, and Figure 9 represents a more generalized version of Figure 8 with instance-agnostic representation.\\ 


\begin{figure*}[ht!]
\begin{adjustwidth}{-2cm}{}
\fbox{\includegraphics[scale=0.47]{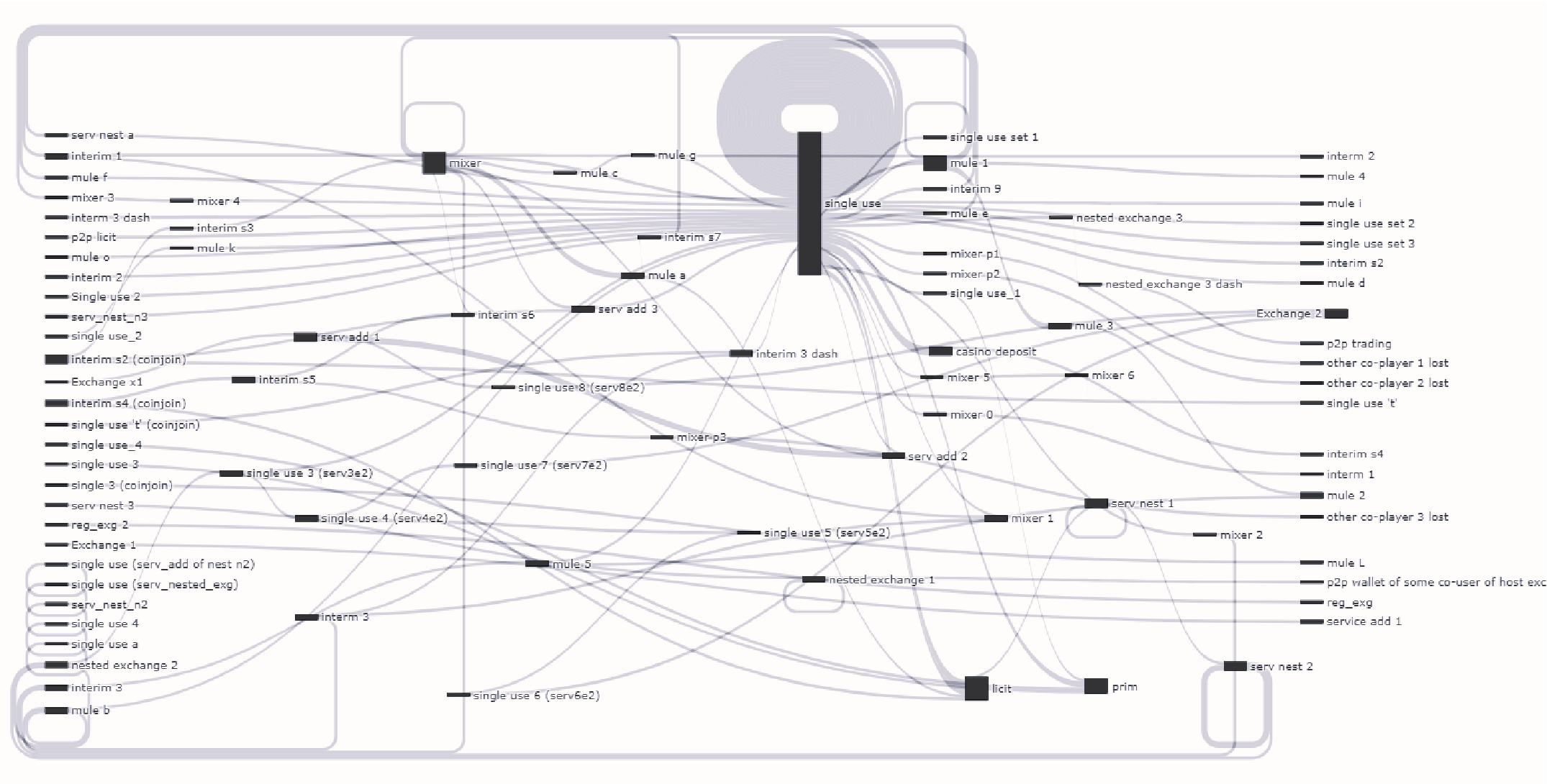}}
\end{adjustwidth}
\caption{Entity-instance-specific alluvial graph}
\label{fig:figure8}
\end{figure*}
\begin{figure*}[ht!]
\fbox{\includegraphics[width=\textwidth]{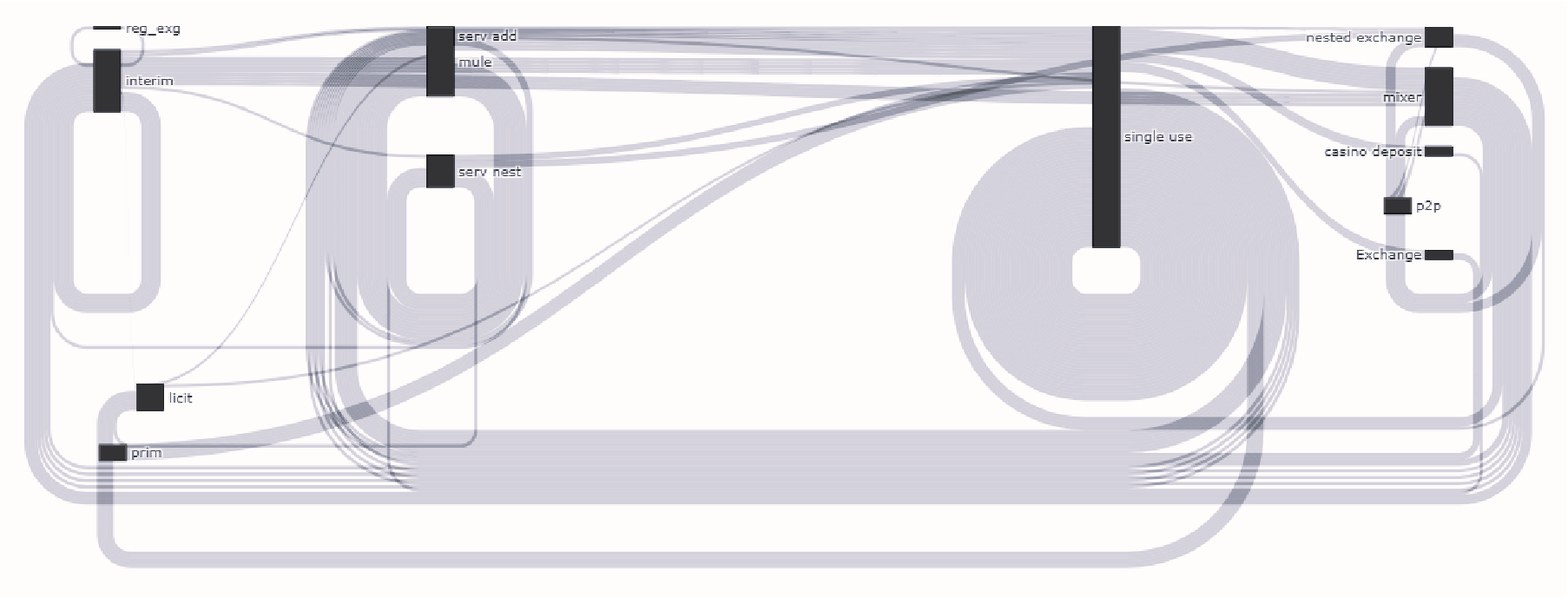}}
\caption{Entity-instance-agnostic alluvial graph}
\label{fig:figure9}
\end{figure*}

\begin{table*}[htbp]
\centering
\begin{adjustbox}{width=\textwidth}
\begin{tabular}{|p{3.5cm}|p{5cm}|p{5cm}|}
 \hline
        \hfil\textbf{Placement methods} & \hfil\textbf{Layering methods} & \hfil\textbf{Integration methods} \\
        \hline
\begin{itemize}[noitemsep,nolistsep]
  \item Deposits in multiple exchanges
  \item Cross-chain conversion (Alt$\leftrightarrow$BTC)
  \end{itemize}
  & \begin{itemize}[noitemsep,nolistsep]
  \item Use of Money mules
  \item Use of Licit accounts, dusting attack
  \item Layered Interim accounts
  \item Cross-chain conversion
  \subitem * Exchange deposits
  \subitem * P2P trading
  \item Hot Car method
  \item Fund splits
  \item Use of Coinjoins
  \item Use of Nested services
  \item Exchange hops
  \item Peel chain
\end{itemize} &
  \begin{itemize}[noitemsep,nolistsep]
  \item Crypto appreciation
  \item Money Service Business (MSB)
  \item Crypto lending
  \subitem{* Centralized lending}
  \subitem{* P2P lending}
  \item Small- and Medium-sized Business (SMB)
\end{itemize}\\
\hline
\end{tabular}
\end{adjustbox}
\caption{Simulatable methods of phases of e2e money laundering}
\label{Table:1}
\end{table*}

A generated transaction has the following attributes:
\begin{enumerate}
\item Unique transaction hash  
\item List of sender accounts  
\item List of recipient accounts   
\item Respective values put in by the sender accounts  
\item Respective values received by the receiver accounts  
\item Timestamp of the transaction  
\item Transaction fees
\end{enumerate}

Figure 10 provides a glimpse of the representation of the generated transactions, and Figure 11 showcases the above-mentioned attributes for a sample transaction.

\begin{figure*}[ht!]
\fbox{\includegraphics[width=\textwidth]{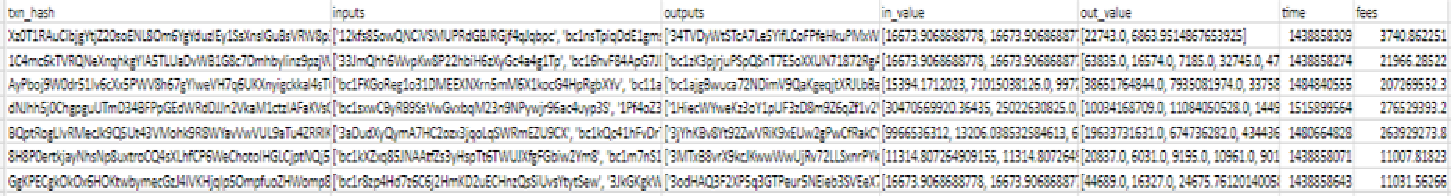}}
\caption{Simulated transactions}
\label{fig:figure10}
\end{figure*}

Using the transactions of an account and its immediate neighbors, we have computed a variety of attributes as discussed further in Section 6 which are fed to the machine learning models to train upon. 

\begin{figure*}[ht!]
\centering
\fbox{\includegraphics[scale=0.75]{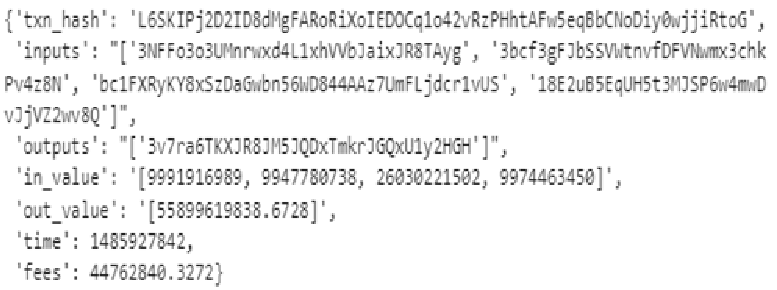}}
\caption{Attributes of a simulated transaction}
\label{fig:figure11}
\end{figure*}

Since we avoided the step of manual data collection and processing, the proposed method required far less hardware than what other available open-source tools need \cite{graphsense}. We were able to create the simulator, generate the data from it, train the machine learning models, and do all the other related activities in a typical machine having just 8 GB of RAM and 256 GB of internal storage with an AMD Ryzen 5 3500U processor. Nevertheless, to speed up the process of generation, at times, we have also used compute queues. However, the environment is not significantly superior. It has 15 GB of RAM and a MIG A100 GPU, which is still far less than what was quoted by other tools.

\section{ML detection tool}
Figure 12 illustrates the architecture of our money laundering detection application along with a utility of the simulator for training machine learning models to identify illicit behavior (money laundering nature). We pass a transaction schema and get the corresponding transactional data generated from the simulator, which we use to train the model. Apart from this data, we may optionally add additional licit or general transactions that are abundantly available for generalization.
\begin{figure*}[ht!]
\fbox{\includegraphics[width=\textwidth]{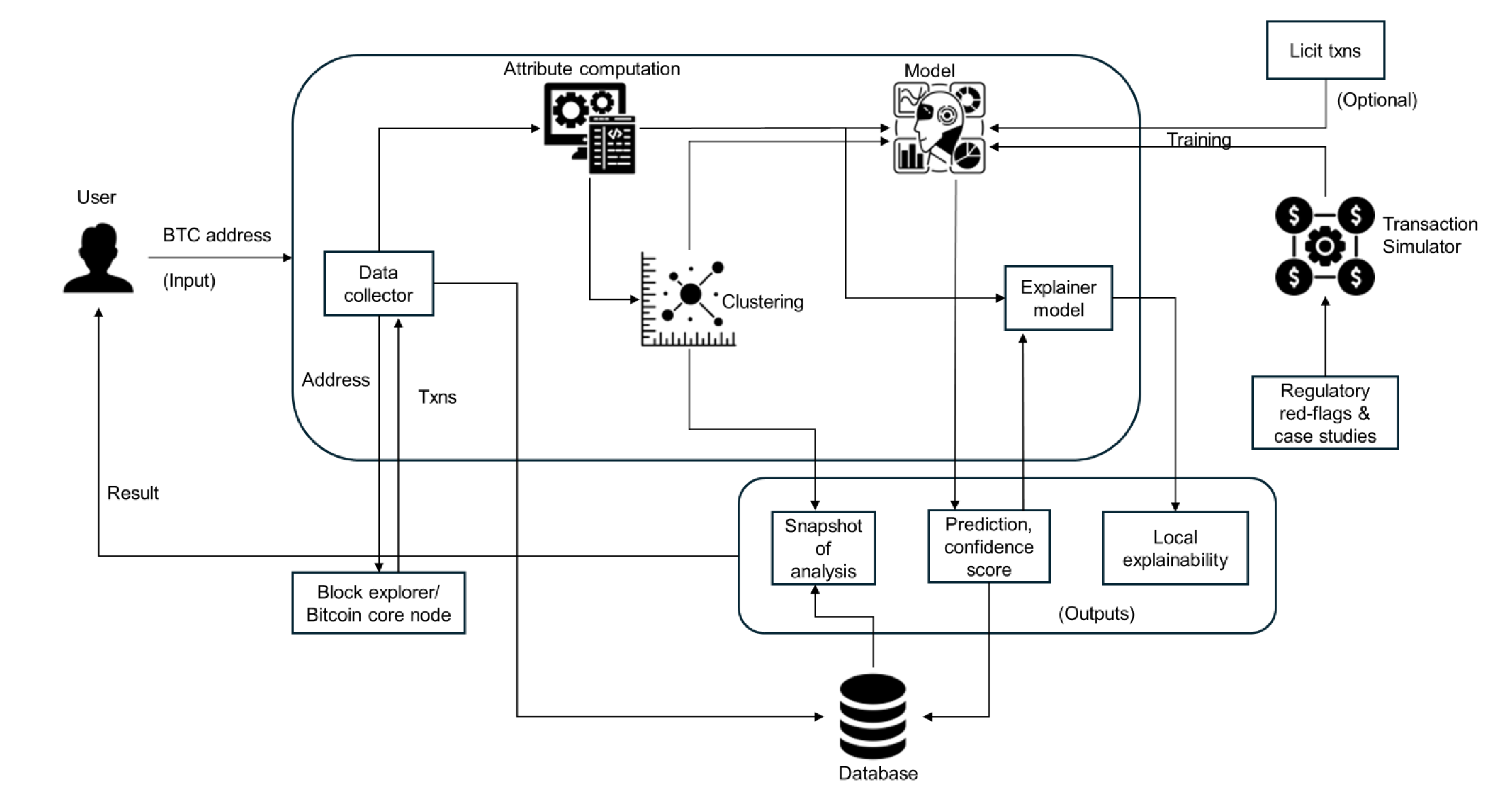}}
\caption{Architecture of ML detection application}
\label{fig:figure12}
\end{figure*}
For using the application, the user would provide a Bitcoin address interested in. Using the data collector, all the real transactions of the provided address will be fetched from public sources such as block explorers, data dumps, etc. Feature engineering is exercised on the collected transactional data for attribute computation where we identify and compute attributes of the data that help in its classification.
\begin{table*}[htbp]
\begin{adjustwidth}{-2.12cm}{}
\centering
\renewcommand{\arraystretch}{1}
\begin{tabular}{|M{1.365cm}|M{0.85cm}|M{0.9cm}|M{1.1cm}|M{1.6cm}|M{0.95cm}|M{0.85cm}|M{1.5672cm}|M{1.941cm}|M{0.85cm}|M{1.5cm}|}
 
 \hline
 \textbf{Model} & \textbf{Train accuracy} & \textbf{Test accuracy} & \textbf{RMSE} & \textbf{Precision} & \textbf{Recall} & \textbf{F1 score} & \textbf{Cross-validation score} & \textbf{Prediction probability} & \textbf{AUC score} & \textbf{Real samples detected (out of 116)}\\
\hline
KNN & 1.0000 & 	0.9996 & 	0.0199 & 	1.0000 & 	0.9992 & 	0.9996 & 	0.9995 & 	0.9995 & 	0.9998 & 	116\\
\hline
Random Forest & 	1.0000 & 	0.9164 & 	0.2891 & 	1.0000 & 	0.8315 & 	0.9080 & 	1.0000 & 	0.7866 & 	0.9988 & 	116\\
\hline
Multilayer Perceptron & 	0.9996 & 	0.9994 & 	0.0244 & 	1.0000 & 	0.9988 & 	0.9994 & 	0.9990 & 	0.9995 & 	1.0000 & 	116\\
\hline
Logistic Regression & 	0.9783 & 	0.9779 & 	0.1485 & 	0.9878 & 	0.9675 & 	0.9775 & 	0.9782 & 	0.9714 & 	0.9907 & 	116\\
\hline
XGBoost & 	1.0000 & 	0.9810 & 	0.1377 & 	0.9640 & 	0.9991 & 	0.9812 & 	1.0000 & 	0.9540 & 	0.9999 & 	4\\
\hline
Decision Tree & 	1.0000 & 	0.8107 & 	0.4350 & 	0.7238 & 	0.9999 & 	0.8398 & 	1.0000 & 	1.0000	 & 0.8123 & 	4\\
\hline
Gradient Boost Classifier & 	0.9998 & 	0.8936 & 	0.3262 & 	0.8234 & 	1.0000 & 	0.9031 & 	0.9997 & 	0.9937 & 	0.9999 & 	4\\
\hline
Naive Bayes & 	0.7684 & 	0.8146 & 	0.4305 & 	0.7972 & 	0.8400 & 	0.8180 & 	0.7684 & 	0.9944 & 	0.8584 & 	116\\
\hline
Ensemble Voting Classifier & 	1.0000 & 	0.9996 & 	0.0209 & 	1.0000 & 	0.9991 & 	0.9996 & 	0.9998 & 	0.9049 & 	1.0000 & 	116\\
\hline
\end{tabular}
\end{adjustwidth}
\caption{Performance metrics of models trained on 70 features}
\label{Table:2}
\end{table*}
As part of this step, we also do clustering to capture additional insights from the data. All the computed attributes are passed to the model which is trained on a similar set of attributes computed from the transactional data generated by the simulator. The trained model would then provide its prediction for the provided Bitcoin address based on the attributes passed to it, created from its transactions. As part of further analytics, we can use an explainer model to understand what attributes contribute to or influence a certain prediction, have a snapshot of analysis showcasing graphical interactions of various addresses seen in the transactional data, risk analytics and so on. All of these results and analytics will be shared with the user.\\
Attributes used to train a machine learning model play a crucial role in defining its spectrum of identifying various money laundering instances. Therefore, it is imperative to identify what attributes we require to identify money laundering that may be done differently by different launderers using different tactics. Thus, to understand this, we have studied several case studies of how money laundering was done in various instances by different people using different approaches and layering methods. We also studied some of the key red-flags and guidelines provided by organizations like FATF \cite{fatf7}, as described in Section 3. Using all this information, which is descriptive, we tried to convert these descriptive patterns into measurable or computable attributes from transactional data of a given account and its neighbors. We also took care that these attributes are not manipulatable from a single transaction point of view. Which is, if I want to make a transaction, I can't tweak it such that it can avoid triggering all the attributes. It would be identified somewhere.\\ 
\begin{table*}[ht!]
\centering
\small 
\begin{tabular}{|p{3cm}|p{10cm}|} 
 \hline
 \hfil\textbf{Model} & \hfil\textbf{Hyperparameters}\\
\hline
KNN & 
n\_neighbors = 7, weights = 'distance', p = 1\\
\hline
Random Forest & n\_estimators = 182, max\_depth = 31, min\_samples\_split = 0.0033588450686818346\\
\hline
Multilayer Perceptron & activation = 'relu', learning\_rate = 'adaptive', max\_iter = 84, alpha = 0.0006892440801331624, early\_stopping = False, solver = 'adam'\\
\hline
Logistic Regression & penalty = 'l2', C = 58.54836598598276, intercept\_scaling = 0.026056129911108616\\
\hline
XGBoost & learning\_rate = 0.030426797451043132, n\_estimators = 157, alpha = 0.01887034997773913, colsample\_bytree = 0.6285193612006074\\
\hline
Decision Tree & max\_depth = 94, min\_samples\_split = 3, criterion = 'gini'\\
\hline
Gradient Boost Classifier & n\_estimators = 63\\
\hline
Naive Bayes & var\_smoothing = 0.001\\
\hline
Ensemble Voting Classifier & ensemble of above methods with soft voting and f1-score as weights\\
\hline
\end{tabular}
\caption{Hyperparameters of the models}
\label{Table:3}
\end{table*}
To maintain the robustness of the features, we exercised the following practices: 
\begin{enumerate}
\item Firstly, features should not be directly manipulatable. 
\begin{enumerate}
\item We considered various aspects of an account for capturing its behavior to identify its class. As discussed, we are translating red-flags and specific patterns into attributes, making them difficult to tweak while making a transaction. The attributes' count comes to be about 130. It includes features based on: primary properties, aggregated features, distribution-based features, time-related features, interaction-based properties and relations, value/amount based features, clustering-based features, frequency-based features, features based on domain factors and characteristics, red-flags, and so on. 
\item If considering these exhaustive set of features while making a transaction, these do not directly map or are manipulatable while making a given transaction.  This is the important shield we put.
\end{enumerate}
\item We use ensemble models so that decision for a sample is not directly based on a linear or non-linear relationship of attributes, or on a specific set of features, which might otherwise be easier to decode. Because, when we use an ensemble of models, each of those models follows different objective functions such as some work based on distance metric, some based on entropy, some based on gradient of the loss function used, some based on similarity metric, and so on. Thus, in this case, it becomes pretty tough to come to a tradeoff with all such different objective functions that work on different principles and still elude. This is another important practice being followed.
\item Use of L1 regularization while training certain models which helps to create sparse models eliminating the significant dependence on one or more features to a greater extent.
\item Adding some noise or multiplying with a scaling factor before training to prevent reverse engineering of identifying the relationship of attributes with one another or with the target variable. This can help in generalization too.
\end{enumerate}

The synthetic dataset having 130 attributes was used to train various models to predict money laundering accounts. The results and performances of the models are presented in Section 7.

\section{Results}
The curated dataset has two target variables: address entity, and address category. However, we focus on the address category which tells us if an account is affiliated with money laundering or not. As discussed in Section 6, to maintain the robustness of the features, and to make the models more generalizable to the variations noticed in real transactions, we have scaled the features with a scaling factor of 1.12 and added a sample-specific noise of $\pm$ 10\% the attribute value to compensate for any 'real' distribution gap.\\
\begin{figure*}[ht!]
\begin{framed}
     \centering
     \begin{subfigure}[ht]{0.49\textwidth}
         \centering
         \includegraphics[width=\textwidth]{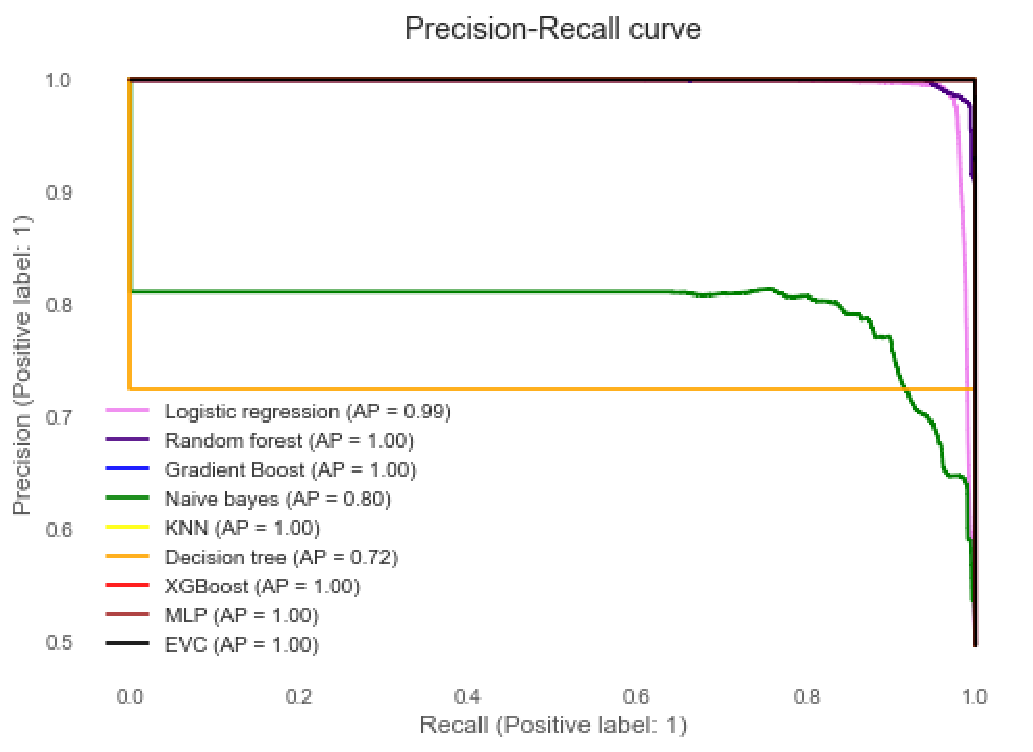}
         \caption{Precision-Recall curve (70 features)}
     \end{subfigure}
     \hfill
     \begin{subfigure}[ht]{0.49\textwidth}
         \centering
         \includegraphics[width=\textwidth]{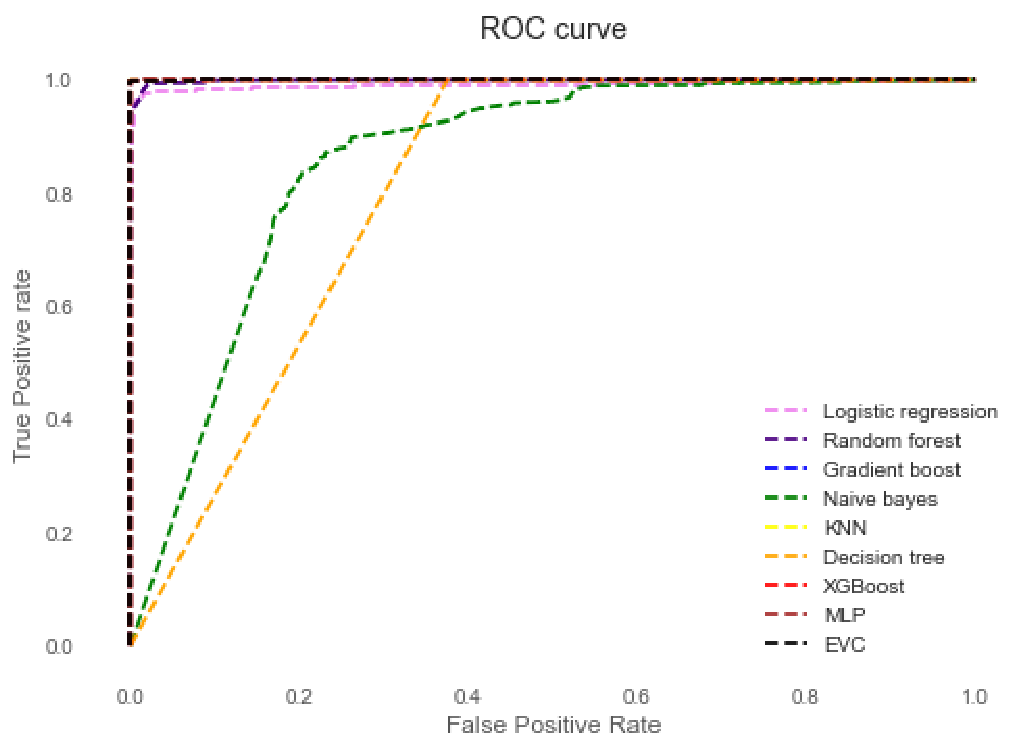}
         \caption{Receiver operating characteristic (ROC) curve (70 features)}
     \end{subfigure}

     \begin{subfigure}[ht]{0.49\textwidth}
         \centering
         \includegraphics[width=\textwidth]{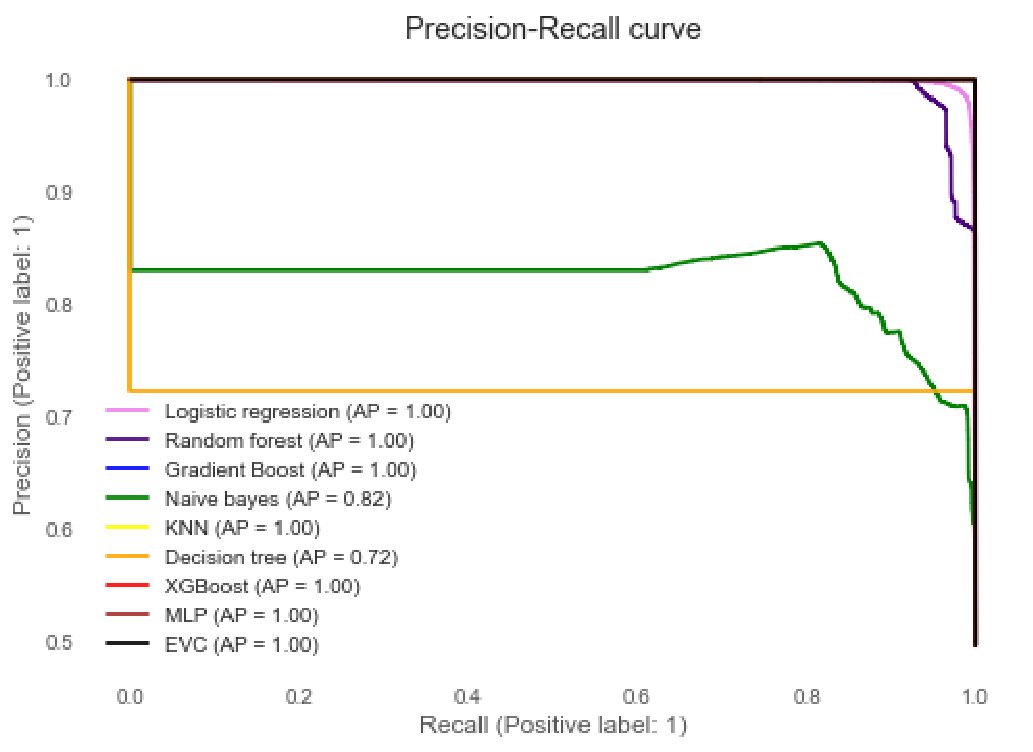}
         \caption{Precision-Recall curve (130 features)}
     \end{subfigure}
     \hfill
     \begin{subfigure}[ht]{0.49\textwidth}
         \centering
         \includegraphics[width=\textwidth]{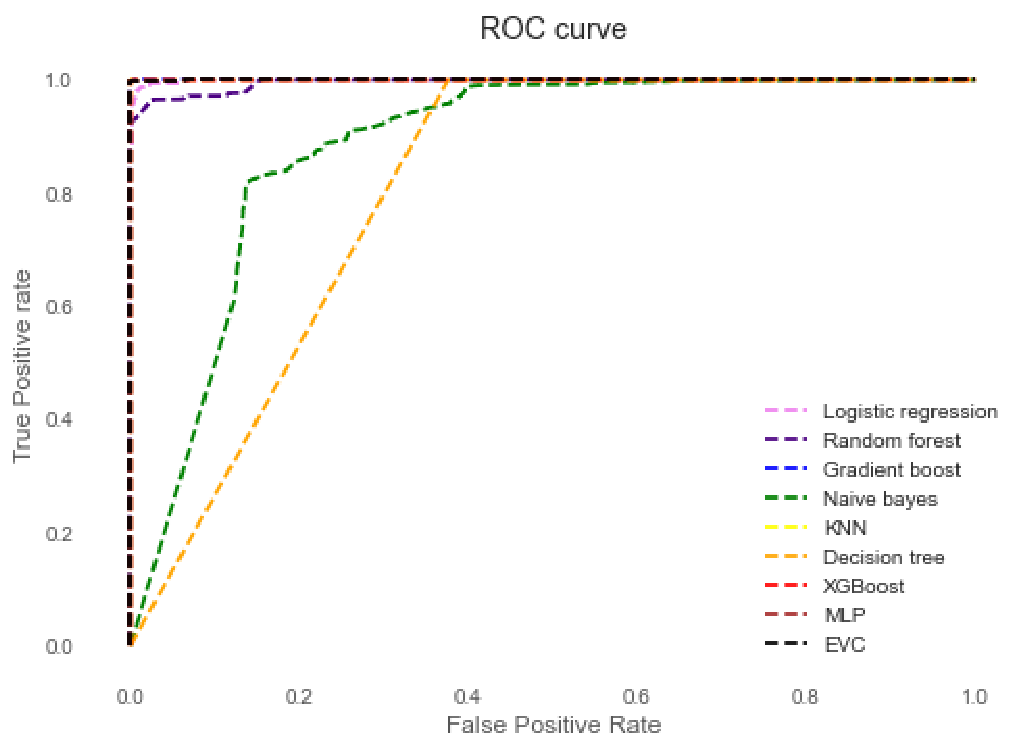}
         \caption{Receiver operating characteristic (ROC) curve (130 features)}
     \end{subfigure}
     \hfill
     \end{framed}
             \caption{Tradeoff and Performance graphs for models trained with 70 and 130 features}
        \label{fig:70ft graphs}
\end{figure*}
We have trained two sets of models upon a partial set of 70 features and the complete set of 130 features respectively. The 70 features are the attributes that can be computed for a real crypto account in real-time, basis its transactional data. The other 60 features are entity interaction-related features that are not available in real-time. These features are captured from the explorative case study conducted as discussed in Section 3, upon deeper insight into their interactional behavior. The same has been replicated in the simulated data using the transaction schema having interactions depicted in Figure 8.\\
\begin{table*}[htbp]
\begin{adjustwidth}{-2.12cm}{}
\renewcommand{\arraystretch}{1}
\centering
\begin{tabular}{ |M{1.4cm}|M{0.85cm}|M{0.85cm}|M{1cm}|M{1.65cm}|M{0.95cm}|M{0.85cm}|M{1.6cm}|M{1.9cm}|M{0.9cm}| } 
 \hline
 \textbf{Model} & \textbf{Train accuracy} & \textbf{Test accuracy} & \textbf{RMSE} & \textbf{Precision} & \textbf{Recall} & \textbf{F1 score} & \textbf{Cross-validation score} & \textbf{Prediction probability} & \textbf{AUC score}\\
\hline
KNN & 
1.0000 & 0.9994 & 0.0252 & 1.0000 & 0.9987 & 0.9994 & 0.9996 & 0.9993 &	0.9998
\\
\hline
Random Forest & 	
1.0000 & 0.8871 & 0.3359 & 1.0000 & 0.7725 & 0.8716 & 1.0000 & 0.7677 & 0.9950
\\
\hline
Multilayer Perceptron & 	
0.9991 & 0.9990 & 0.0315 & 0.9998 & 0.9982 & 0.9990 & 0.9994 & 0.9987 & 0.9999
\\
\hline
Logistic Regression & 	
0.9870 & 0.9855 & 0.1203 & 0.9917 & 0.9790 & 0.9853 & 0.9865 & 0.9891 & 0.9983
\\
\hline
XGBoost & 	
1.0000 & 0.9996 & 0.0209 & 1.0000 & 0.9991 & 0.9996 & 1.0000 & 0.9788 & 1.0000
\\
\hline
Decision Tree & 	
1.0000 & 0.8104 & 0.4355 & 0.7234 & 1.0000 & 0.8395 & 1.0000 & 1.0000 & 0.8119
\\
\hline
Gradient Boost Classifier & 	
0.9999 & 0.9996 & 0.0209 & 1.0000 & 0.9991 & 0.9996 & 0.9997 & 0.9979 & 1.0000
\\
\hline
Naive Bayes & 	
0.7660 & 0.8064 & 0.4400 & 0.8466 & 0.7446 & 0.7923 & 0.7652 & 0.9931 & 0.8827
\\
\hline
Ensemble Voting Classifier & 	
0.9999 & 0.9995 & 0.0227 & 1.0000 & 0.9990 & 0.9995 & 0.9999 & 0.9226 & 1.0000
\\
\hline
\end{tabular}
\end{adjustwidth}
\caption{Performance metrics of models trained on 130 features}
\label{Table:4}
\end{table*}
Table 2 shows the performance metrics of various models trained upon the set of 70 features. The last column in the table(-2) is the number of real illicit accounts the respective model was able to identify based on training only upon the simulated data. For this, we have used the list of 116 real illicit (Bitcoin) accounts \cite{a116}, as published by OFAC in its sanction reports, involved in money laundering by the Lazarus group. Many of the finetuned models upon the respective hyperparameters mentioned in Table 3 were able to identify all of the real money laundering accounts.\\
Figures 13 (a), 13(b) are the precision-recall curve and receiver operating characteristic (ROC) curve for the models trained on 70 features. The same for the models trained on 130 features is represented in Figures 13(c) and 13(d).\\
When trained upon all the 130 attributes, the performance of Gradient Boost Classifier was significantly improved, considering the metrics of test accuracy and F1-score, and there has been a slight improvement in XGBoost and Logistic Regression. While there is a slight drop in Random Forest and Naive Bayes, the remaining models remained almost the same, as demonstrated in Table 4. This observation is noted when we used the same hyperparameters for both the sets of features.
\\
\begin{figure*}[ht!]
\fbox{\includegraphics[width=\textwidth]{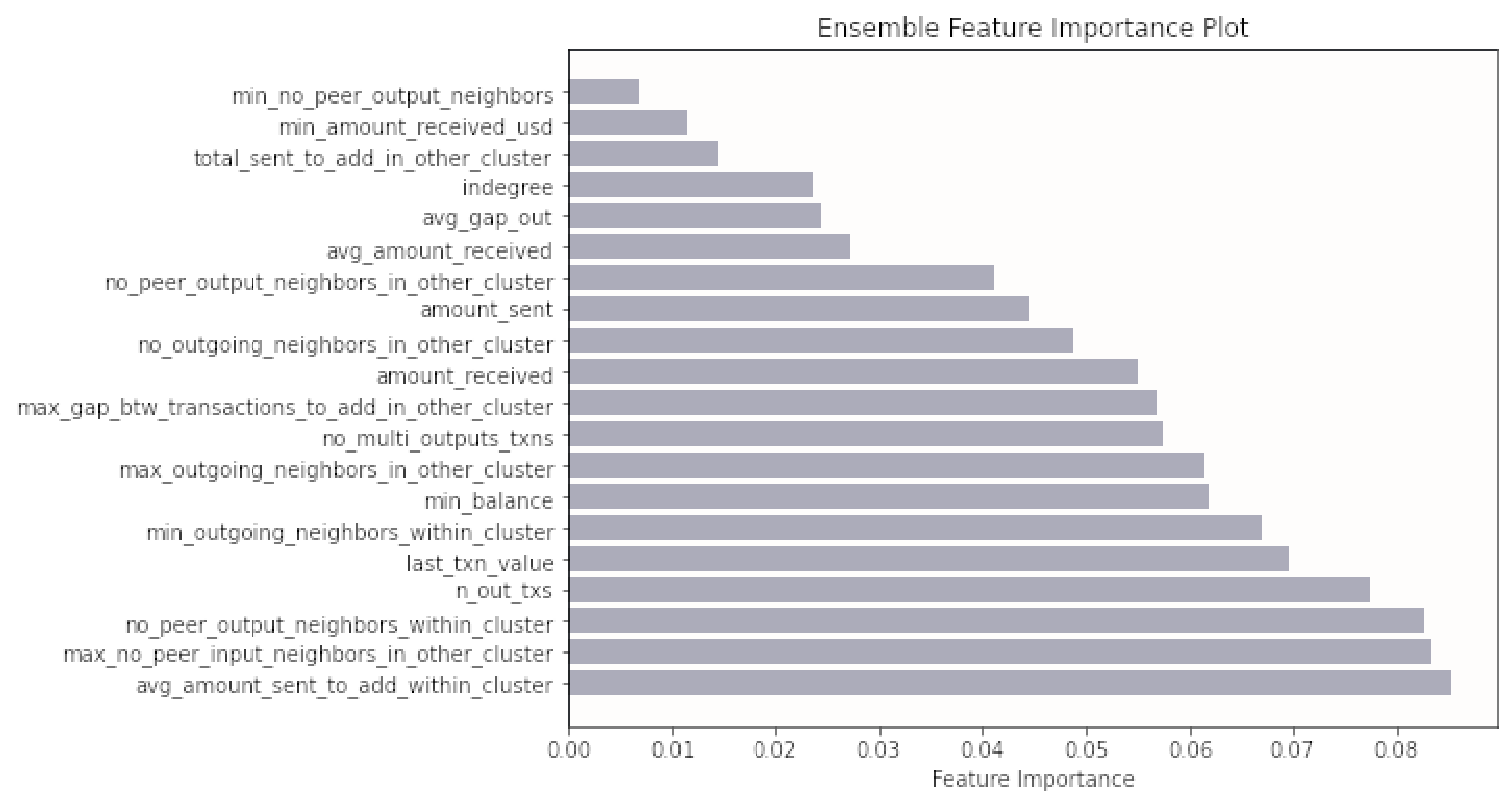}}
\caption{Ensemble feature importance}
\label{fig:figure14}
\end{figure*}
Figure 14 is the plot of important features obtained through an ensemble of three feature importance measuring techniques namely information gain, feature shuffling, and feature performance.\\

\section{Conclusion and Future Directions}
Money laundering in cryptocurrency has been a pertinent problem predominantly becoming a way to move funds from all sorts of illicit activities into a legal financial ecosystem, causing financial loss and reputational damage to people, organizations and so on apart from many other negative effects. This has been due to many special characteristics of cryptocurrencies. Addressing crypto money laundering also shows severe impact on the diminishment of many other illicit activities ranging from small-scale scams to large-scale terrorism financing. However, crypto money laundering detection is pretty hard due to a variety of hardships ranging from data collection, verification to handling, and processing it amongst many other factors as discussed in the paper.\\
To address this very primary challenge, we have developed a behavior-embedded entity-specific money laundering-like transaction simulator. This facilitates in the generation of custom transactions and datasets corresponding to the usecase or requirement from a simple transaction schema. The paper discusses the design and architecture of the simulator, along, with the custom dataset we created, and the performance of the machine learning models trained upon it in detecting real accounts involved in money laundering.\\
We believe, this work paves the way for advanced data generation techniques, such as training sophisticated generative adversarial networks (GANs) using the custom simulated data from the simulator, for better generalization of patterns and effective detection of illicit accounts and so on.

\section*{Statements and Declarations}
\textbf{Author contributions} \textit{Dinesh} has made the explorative study, designed and implemented the simulator and tool, executed the models, prepared the figures, and wrote the paper.
\textit{Manoj} provided domain insights, guidance and course correction of the work, verified and validated different phases of the work including the results, and reviewed the paper.\\

\hspace{-0.6cm}
\textbf{Funding} The authors work at the affiliated institution and they have not received any special funding for this work.\\

\hspace{-0.6cm}
\textbf{Conflicts of interest} The authors have no competing interests to declare that are relevant to the content of this article and agree to the
publishing of its content.\\

\hspace{-0.6cm}
\textbf{Ethics approval} Not applicable.\\

\hspace{-0.6cm}
\textbf{Data availability} The data associated with this research paper is not openly available due to two primary reasons. Firstly, it constitutes the intellectual property of the authors' institution. Secondly, the dataset contains attributes carefully designed from an extensive exploratory study to identify money laundering, which, if released openly, poses a significant risk of misuse. However, researchers with genuine interest in accessing and experimenting with the data may contact the authors, specifying their purpose of use. Depending on the nature of the request and adherence to usage-specific terms and conditions, the authors may consider sharing the data accordingly. Further, the models experimented with in this study, along with their corresponding hyperparameters, are specified in the paper, and no external data was used to train or fine-tune the specified models.\\

\bibliography{refs}%
\end{document}